\documentclass[aps,prb,twocolumn,showpacs,citeautoscript,longbibliography,superscriptaddress]{revtex4-1}

\usepackage{amsmath,amssymb}
\usepackage{bm}
\usepackage{graphicx}
\usepackage{epstopdf} 
\usepackage{latexsym} 
\usepackage{subfigure}
\usepackage{color}
\usepackage{dsfont} 
\usepackage{wasysym}

\begin{document} 
\title{Electron quasi-itinerancy intertwined with quantum order by disorder 
\\
in pyrochlore iridate magnetism}
\author{Gang Chen}
\affiliation{Department of Physics and HKU-UCAS Joint Institute 
for Theoretical and Computational Physics at Hong Kong, 
The University of Hong Kong, Hong Kong, China}
\affiliation{State Key Laboratory of Surface Physics and Department of Physics, 
Institute of Nanoelectronics and Quantum Computing, 
Fudan University, Shanghai 200433, China}
\affiliation{Collaborative Innovation Center of Advanced Microstructures, 
Nanjing University, Nanjing 210093, China}
\author{Xiaoqun Wang}
\affiliation{School of Physics and Astronomy, Tsung-Dao Lee Institute, 
Shanghai Jiao Tong University, Shanghai 200240, China}
\affiliation{Key Laboratory of Artificial Structures and Quantum Control of MOE,
Shenyang National Laboratory for Materials Science, Shenyang 110016, China}

\date{\today}

\begin{abstract}
We point out the emergence of magnetism from the interplay of electron quasi-itinerancy
and quantum order by disorder in pyrochlore iridates. Like other Mott insulating iridates, 
the Ir$^{4+}$ ion in pyrochlore iridates develops an effective ${J=1/2}$  
moment from the on-site spin-orbit coupling. We consider the generic symmetry-allowed 
exchange between these local moments on a pyrochlore lattice and obtain the mean-field 
phase diagram. Assuming the superexchange is mediated by direct and/or indirect electron
hopping via intermediate oxygens, we derive the exchange interactions in the strong  
coupling regime from the Hubbard model. This exchange has a degenerate classical 
ground state manifold and quantum fluctuation selects a non-coplanar ground state, 
known as quantum order by disorder. Extending to the intermediate coupling regime, 
the same non-coplanar order is selected from the degenerate manifold by the kinetic energy, 
which is dubbed ``electron quasi-itinerancy''. We discuss the experimental relevance of 
our results and electron quasi-itinerancy among other iridates and $4d/5d$ magnets. 
\end{abstract}

\date{\today}

\maketitle

\section{Introduction}
\label{sec1}

In recent years, there have been a lot of activities on the Ir-based transition metal oxides. 
Due to the strong spin-orbit coupling (SOC) of its $5d$ electrons, many novel phases, 
theoretical models and experiments have been proposed and discovered in 
these Ir-based materials~\cite{2014ARCMP...5...57W,2016ARCMP...7..195R,2019arXiv190308081T,Schaffer_2016}. 
Among them, for example, a quantum spin liquid phase was proposed for 
an Ir-based hyperkagom\'{e} lattice in Na$_4$Ir$_3$O$_8$~\cite{PhysRevLett.99.137207}; 
a ferromagnetic ground state with a large ferromagnetic moment was discovered 
in Sr$_2$IrO$_4$ with the Ir$^{4+}$ ions forming a square lattice~\cite{PhysRevB.57.R11039}. 
In these Mott insulating systems, the presence of strong SOC drastically 
changes the local spin physics.  The local moment of the magnetic ion 
Ir$^{4+}$ is an effective ${J = 1/2}$ moment~\cite{PhysRevLett.99.137207,hyperk_chen,PhysRevLett.102.017205}
describing local spin-orbital doublets rather than usual electron spin ${S = 1/2}$ 
for systems with a weak SOC. The existence of local spin-orbital doublets has been 
detected by resonant X-ray scattering experiment in Sr$_2$IrO$_4$~\cite{xray}. 
As a consequence, the non-trivial exchange interaction can arise due to the mixing 
of spin and orbitals~\cite{PhysRevLett.99.137207,PhysRevLett.105.027204}.

Even though the superexchange interaction between the Ir local moments is often used 
to describe most iridates, most well-known Mott insulating iridates are actually weak 
Mott insulators with quasi-itinerant $5d$ electrons and small charge gaps. 
This weak Mott insulating nature was not really emphasized in the literature, 
and we think this may be important in understanding some of the physical 
properties of iridates and the related materials. What is electron quasi-itinerancy? 
Quasi-itinerancy is the key property of the electrons in the weak Mott regime where 
the Mott gap is not large enough to fully localize the electron to one single lattice site, 
and the electron can still be delocalized to a finite extent spatially due to the small charge gap. 
Electron quasi-itinerancy is believed to the driving force for the possible spin liquid phase 
in the weak Mott regime for $\kappa$-(ET)$_2$Cu$_2$(CN)$_3$ and 
EtMe$_3$Sb[Pd(dmit)$_2$]$_2$~\cite{PhysRevLett.95.036403}. 
Over there, the electron quasi-itinerancy
generates the frustrated ring exchange interactions that suppresses the magnetic orders. 
Similar kind of electron quasi-itinerancy~\cite{PhysRevLett.108.247215,PhysRevLett.107.186403,PhysRevLett.94.156402} 
that emphasizes different outcomes of the charge fluctuations 
has been discussed in various spinels and osmate pyrochlores. 
Thus, besides the prevailing strong coupling perspectives, the weak to intermediate coupling
perspective is found to be both complementary and exciting. Ref.~\onlinecite{PesinBalents} 
applied a slave-rotor mean field theory to study the Mott transition in a series of rare-earth 
based pyrochlore iridates R$_2$Ir$_2$O$_7$. They discovered topological band insulator
in the non-interacting limit and a novel topological Mott insulator in intermediate coupling regime. 
Several other groups re-examined the problem with more realistic Hamiltonian and discovered 
various magnetic ordered phases and an interesting Weyl semi-metal phase that is located 
in the narrow regime separating the topological band insulator or metal phase from strong 
coupling Mott insulating phase~\cite{PhysRevLett.109.066401,PhysRevB.85.045124,PhysRevB.83.205101}. 
Aligned with the above theoretical efforts, the experiments discovered that a metal-insulator
transition in R$_2$Ir$_2$O$_7$ (R = Nd, Sm, Y and Eu) involves a magnetic ordering produced 
by the $5d$ electrons in Ir~\cite{PhysRevB.85.245109,PhysRevLett.109.136402,PhysRevB.85.214434,PhysRevB.86.014428,PhysRevB.85.205104,PhysRevB.87.100403,PhysRevLett.117.037201,PhysRevLett.115.056402,PhysRevB.89.140413,PhysRevB.89.115111,PhysRevLett.117.056403,PhysRevB.88.060411,PhysRevB.89.075127,PhysRevLett.114.247202,PhysRevB.87.060403,PhysRevB.90.235110,PhysRevB.83.180402,PhysRevB.94.161102,PhysRevB.92.094405,PhysRevB.96.094437,PhysRevMaterials.2.011402,2020arXiv200512768W,PhysRevB.101.121101,PhysRevB.96.144415}. Moreover, an exotic spin liquid metallic phase was also proposed 
experimentally for Pr$_2$Ir$_2$O$_7$~\cite{PhysRevLett.96.087204,Machida}. 
Now, the Ir electrons are proposed as Luttinger 
semimetal~\cite{2015NatCo...610042K,PhysRevX.4.041027,PhysRevLett.111.206401},
while the Pr spin is proposed to be proximate to a transition between a U(1) spin liquid 
and an ordered spin ice~\cite{PhysRevX.8.041039,PhysRevB.94.205107,2017arXiv171107813O,PhysRevB.92.054432}.

Based on the existing theoretical~\cite{PesinBalents,kim_distortion,PhysRevB.85.045124,PhysRevB.86.235129,PhysRevB.96.195158,PhysRevB.91.115124,PhysRevX.8.041039,PhysRevLett.112.246402,PhysRevLett.109.066401,PhysRevB.87.214416,PhysRevLett.111.206401,PhysRevB.95.045133,PhysRevX.4.041027} and experimental works,
the true magnetic state of these Ir-based pyrochlore
systems remains open. In the present paper, we 
address this problem and provide some understanding. 
We primarily focus on the magnetic properties and avoid
touching the band structure topology that has been 
invoked in early works. We first explore the magnetic properties of 
the Ir-based pyrochlore lattice in the strong coupling regime. 
Physically, since the $5d$ electron orbitals of the Ir$^{4+}$ 
is spatially extended, which enhances the electron bandwidth, 
therefore these Ir-based systems are usually considered to be 
in the intermediate coupling regime. Nevertheless, 
the SOC could enhance the correlation by suppressing the 
bandwidth~\cite{PesinBalents}. 
Moreover, certain magnetic 
properties in the strong coupling limit could persist to the intermediate
coupling regime even if the system is located in the intermediate coupling regime. 
In the strong coupling limit, the effective moments 
${J = 1/2}$ of the Ir$^{4+}$ ions are coupled by the superexchange
interaction. We analyze the symmetry-allowed exchange
Hamiltonian, that includes three types of pairwise terms, 
Heisenberg exchange, antisymmetric Dzyaloshinskii-Moriya (DM) 
interaction and the symmetric pseudo-dipolar (PD) interaction.    
This model is equivalent to the one that was used for 
interacting Kramers doublet for the rare-earth 
pyrochlores. 
In the mean-field phase diagram, we find five different ordered phases 
(see Sec.~\ref{sec2}): 4-in-4-out state, a continuously degenerate state 
spanned by two basis vectors ${({\bf v}_1, {\bf v}_2)}$, a weakly ferromagnetic state 
(FM) and two coplanar states with spin orient along the particular $[110]$
directions. Almost all these ordered states have the magnetic 
wavevector ${{\bf q} = 0}$. For the realistic exchange model 
obtained from an extended Hubbard model relevant for R$_2$Ir$_2$O$_7$,
there are only two ordered phases, that are the “4-in-4-out”
state and the continuously degenerate manifold spanned by two
basis vectors ${({\bf v}_1, {\bf v}_2)}$. For the latter, we find that 
the quantum fluctuation selects a non-coplanar spin configuration 
by a linear spin-wave expansion. This is the mechanism of quantum 
order by disorder. For the intermediate coupling regime, we apply the 
self-consistent mean-field theory to study the microscopic Hubbard model 
and assume a general magnetic configuration except having the same  
magnetic cell as the crystallographic cell (or ${{\bf q} = 0}$) order. 
Again, we find the system is ``fluctuating'' within the continuously 
degenerate manifold spanned by ${({\bf v}_1, {\bf v}_2)}$, and 
the electron kinetic energy selects the magnetic orders. The electron 
kinetic energy is important here due to the quasi-itinerancy in the weak 
Mott regime. It is found that the magnetic orders in the strong coupling 
regime persist into the intermediate coupling regime. 
Since it is unclear which regime the actual system is proximate to, 
it is reasonable to think the electron quasi-itinerancy is intertwined
with the quantum order by disorder here.

In the following, we outline the main content 
of this paper. In Sec.~\ref{sec2}, we study a generic symmetry-allowed 
exchange Hamiltonian on the pyrochlore lattice with the effective spin-1/2 originating 
from Kramers' degeneracy, that is relevant for R$_2$Ir$_2$O$_7$ in the strong 
coupling regime. In the exchange Hamiltonian, there are four symmetry allowed
coupling parameters, Heisenberg exchange $J_0$, DM interaction $D$, and
$\Gamma_1$, $\Gamma_2$ for PD interaction. We analyze this Hamiltonian 
with the mean-field method in different parameter regimes. In many
parts of phase diagram, the ground state can be understood as
simultaneously optimizing different terms of the Hamiltonian.
In Sec.~\ref{sec3}, we derive a realistic exchange from the
extended Hubbard model. Two limits with the dominant direct or indirect 
electron tunneling via intermediate oxygens are considered. In these
two cases, we find there is only one mean-field phase, which is
the continuously degenerate manifold ${({\bf v}_1, {\bf v}_2)}$. 
We then implement the linear spin-wave theory and a non-coplanar 
ground state is favored by this quantum order by disorder mechanism. 
For the certain intermediate regime with comparable direct and 
indirect electron tunnelings, the “4-in-4-out” state is favored. 
We further explore the magnetic properties of the Hubbard model 
in the intermediate coupling regime. By assuming a ${{\bf q} = 0}$ 
magnetic structure, we implement a Hartree-Fock
type of self-consistent mean-field theory for the interaction.
Finally in Sec.~\ref{sec4}, we discuss the relevant experiments and 
other related works.

\section{The generic exchange Hamiltonian}
\label{sec2}

In this section, we analyze the Ir-based pyrochlore lattice in
strong coupling regime. In the strong coupling limit, the local
effective spin moments are coupled by an exchange Hamiltonian. 
For the effective spin-1/2 moment describing the local Kramers' 
doublets, the exchange interaction is guaranteed to
be pairwise. The generic exchange Hamiltonian has the following form
\begin{eqnarray}
{\mathcal H}_{\text{ex}} &=& \sum_{\langle ij \rangle} J_0 ( {\bf J}_i \cdot {\bf J}_j )
+ {\bf D}_{ij} \cdot ( {\bf J}_i \times {\bf J}_j )
 + { {\Gamma}}^{\mu\nu}_{ij} J^{\mu}_i J^{\nu}_j,
\label{eq:exchange}
\end{eqnarray}
where the nearest-neighbor interaction is assumed,
and $J_0$ is the isotropic Heisenberg exchange, ${\bf D}_{ij} $ describes
the antisymmetric Dzyaloshinskii-Moriya (DM) interaction
and $ {\Gamma}_{ij}^{\mu\nu}$ is the symmetric pseudo-dipolar (PD) interaction. 
This form of decomposition is well-known to the much older 
literature of magnetism~\cite{tmo}, but is not quite popular among newer ones. 
Kitaev or any other anisotropic exchange interactions can be well cast into this form, 
as long as they are {\sl pairwise} interactions. As a general rule of thumb, 
for systems with a weak SOC, the DM interaction is weaker than
to the Heisenberg part, and PD interaction is even weaker than DM interaction. 
For systems with strong SOC such as iridates here, there is no general thumb of rule,
and all the interactions could be of similar magnitudes. Thus, for most magnetic 
systems composed of $3d$ transition metal ions, 
the DM interaction and PD interaction are expected to be much weaker than 
the Heisenberg exchange and hence can be neglected at lowest order approximation. 
For Ir-based magnets or other magnetic systems formed by $4d$/$5d$ transition metal ions, 
SOC is quite strong and local moment is a mixture of spin and orbitals. As a result, 
the exchange interaction is usually very non-Heisenberg-like and the anisotropic 
exchanges (such as DM and PD interactions) can be quite significant.

\begin{figure}[t]
\includegraphics[width=8cm]{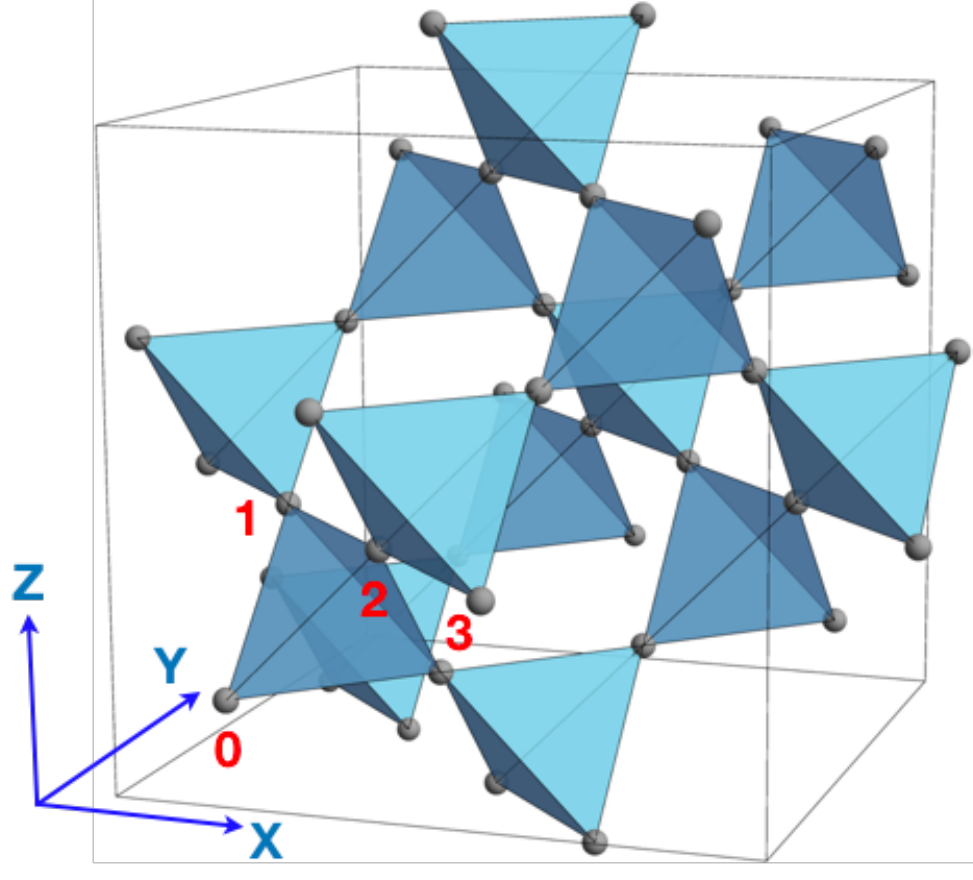}
\caption{(Color online) The pyrochlore lattice in the global cubic coordinate system. 
``0,1,2,3'' label the four sublattices.} 
\label{fig1}
\end{figure}

Throughout this section, we assume an antiferromagnetic Heisenberg part 
with ${J_0 >0}$. Since most R$_2$Ir$_2$O$_7$ (and also spinel AB$_2$X$_4$) 
compounds have a space group Fd$\bar{3}$m, this space group symmetry 
further restricts the allowed forms of the DM interaction and PD interaction. 
Therefore, for the bond connecting the sublattice 0 with the sublattice 1 
(see Fig.~\ref{fig1}), we have,
\begin{eqnarray}
{\bf D}_{01} &=& D (0,\frac{1}{\sqrt{2}},-\frac{1}{\sqrt{2}}),
\\
\Gamma_{01} &=&
\left[
\begin{array}{ccc}
-2\Gamma_1 & 0 & 0 \\
0           & \Gamma_1 & -\Gamma_2 \\
0           & -\Gamma_2 & \Gamma_1 
\end{array}
\right],
\end{eqnarray} 
where the matrix $\Gamma_{01}$ is demanded to be symmetric and traceless as the traceful part
of the full interaction is taken care of by the Heisenberg interaction, and anti-symmetric one is from
the DM interaction. Exchange interactions on other bonds can be simply generated by cubic permutations.

\begin{figure*}[t]
\includegraphics[width=18cm]{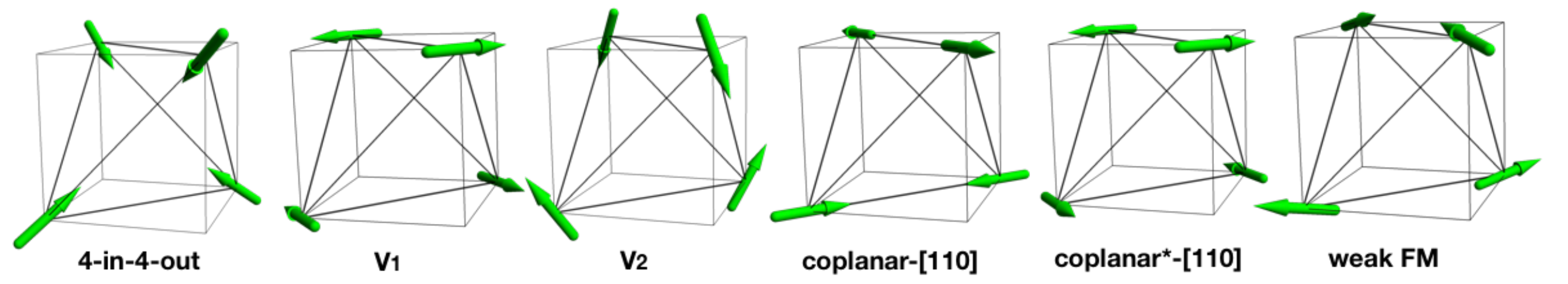}
\caption{(Color online.) The spin configuration on each sublattice for different phases.}
\label{fig2}
\end{figure*}

Although the exchange Hamiltonian in Eq.~\eqref{eq:exchange} is introduced for Ir-based 
pyrochlore lattice, it is widely applicable to other pyrochlore systems with the same symmetry 
properties as long as the local moment is a Kramer spin-1/2 doublet. Our results would also apply
to these contexts as well. In fact, this model is equivalent to the one that was used for the 
rare-earth pyrochlore material 
Yb$_2$Ti$_2$O$_7$ where some detailed analysis were given in Ref.~\onlinecite{PhysRevX.1.021002,PhysRevB.95.094422}. 
Over there, a local coordinate system was 
used for each pyrochlore sublattice and the local moment is the Kramers doublet of the Yb$^{3+}$ 
ion, while here we are using a global cubic coordinate for the Ir$^{4+}$ effective spin-1/2 moments. 
In the next subsection, we analyze the mean-field ground states of this general Hamiltonian 
and understand the role of different anisotropic interactions.

\subsection{Role of Dzyaloshinskii-Moriya interaction}  

Here we consider the role of Dzyaloshinskii-Moriya interaction on top of the 
Heisenberg interaction and set ${ \Gamma_1=\Gamma_2 = 0}$. 
Classically, it is well-known that the pyrochlore lattice is the most frustrated lattice 
by having a macroscopic number of ground state degeneracies with the nearest-neighbor 
Heisenberg model. The presence of the anisotropic exchange surely lifts the classical 
ground state degeneracy. Ref.~\onlinecite{pyrochlore_DM} has already studied 
the role of DM interaction by mean-field theory and classical Monte-Carlo simulation. 
Our mean-field analysis below by treating the effective spin ${\boldsymbol J}_i$
as a classical vector is consistent with their results. With a direct DM interaction   
that corresponds to ${D<0}$ in the present work, the ground state is 2-fold degenerate 
(related by time reversal) with the magnetic ordering wavevector ${\bf q} = {\bf 0}$.  
The magnetic unit cell coincides with the crystallographic one and the four spins
on the unit cell are
\begin{eqnarray}
\Psi &\equiv& ({\bf J}_0, {\bf J}_1, {\bf J}_2, {\bf J}_3 ) 
\nonumber \\
&=& \frac{1}{\sqrt{3}} (111,1\bar{1}\bar{1},\bar{1}1\bar{1}, \bar{1}\bar{1}1).
\end{eqnarray}
Here we define a vector $\Psi$ for the four spin vectors on the elementary tetrahedron. 
This is the simple 4-in-4-out state. 

For the indirect DM interaction with ${D>0}$, DM interaction only partially 
lifts the ground state degeneracy. There are two sets of ground states, 
coplanar and non-coplanar states, both of which have a magnetic wavevector 
${{\bf q} = {\bf 0} }$.  The 4-spin vector $\Psi$ of the coplanar ground states can 
be constructed as linear superpositions of the following two basis vectors 
${\bf u}_1$ and ${\bf u}_2$
(or their equivalence under discrete symmetry operations)
\begin{eqnarray}
{\bf u}_1&=& (100,010,0\bar{1}0,\bar{1}00)  ,
\label{eq:basiscp1}
\\
{\bf u}_2 &=& (010,\bar{1}00,100,0\bar{1}0).
\label{eq:basiscp2}
\end{eqnarray}
The non-coplanar states are constructed from the 
following two basis vectors ${\bf v}_1$ and ${\bf v}_2$
(or their symmetry equivalence)
\begin{eqnarray}
{\bf v}_1 &=& \frac{1}{ \sqrt{2} } (\bar{1}10,\bar{1}\bar{1}0,110,1\bar{1}0) ,
\label{eq:basis1_dm}
\\
{\bf v}_2 &=& \frac{1}{\sqrt{6}  } (\bar{1}\bar{1}2,\bar{1}1\bar{2},1\bar{1}\bar{2},112).
\label{eq:basis2_dm}
\end{eqnarray}
Here, when only the first basis vector ${\bf v}_1$ is chosen, the ground state 
is a special coplanar state with spin orient along different $[110]$ lattice directions. 
Both the coplanar and non-coplanar degenerate ground state manifolds have an
accidental U$(1)$ degeneracy with one continuous degree of freedom.
This degenerate spin manifold is actually identical to the one 
that was proposed for the rare-earth pyrochlore Er$_2$Ti$_2$O$_7$, and
are selected via the quantum order by disorder mechanism~\cite{PhysRevLett.109.167201}.

\subsection{Role of pseudo-dipolar interaction: case 1}

In this and next subsections, we study the role of the PD interaction. 
We first consider the regime with ${D=0}$, ${\Gamma_1 \neq 0}, {\Gamma_2 =0}$. 
For ${\Gamma_1>0}$, we find that optimal spin configurations have 
magnetic wavevector ${{\bf q}={\bf 0}}$. Even though the Hamiltonian 
breaks the spin rotation symmetry completely, the ground state     
manifold has an accidental O$(3)$ degeneracy. The 4-spin vector $\Psi$ 
of the ground states is an arbitrary linear superposition of the 
following three basis vectors ${\bf w}_1$, ${\bf w}_2$ and ${\bf w}_3$,
\begin{eqnarray}
{\bf w}_1 &=&(100,100,\bar{1}00,\bar{1}00),   \\
{\bf w}_2 &=&(010,0\bar{1}0,010,0\bar{1}0),   \\
{\bf w}_3 &=& (001, 00\bar{1}, 00\bar{1},001).
\end{eqnarray}

For ${\Gamma_1 <0}$, to simultaneously optimize the energy and 
satisfy the hard spin constraint, there only exist two 
sets of ground states. One has the magnetic wavevector ${\bf q} = {\bf 0}$. 
Similar to the case with ${\Gamma_1 >0}$, 
the ground state spin configuration has O$(3)$ degeneracy 
and the 4-spin vector $\Psi$ is constructed from the following 
three basis vectors ${\bf z}_1, {\bf z}_2$ and ${\bf z}_3$ 
(or their symmetry equivalence),  
\begin{eqnarray}
{\bf z}_1 &=& ( 100, \bar{1}00, \bar{1}00, 100 ), 
\\
{\bf z}_2 &=& ( 0{1}0, 0\bar{1}0, 0\bar{1}0, 0{1}0 ), 
\\
{\bf z}_3 &=& ( 001, 001, 00\bar{1}, 00\bar{1} ).
\end{eqnarray}

The other set of ground states has the magnetic wavevector ${{\bf q} = 2\pi(100)}$ 
or its cubic equivalences. Although the magnetic unit cell doubles the size of 
crystallographic cell, the spin configuration can still be fully described within 
one tetrahedron and the 4-spin vector $\Psi$ is given as
\begin{equation}
\Psi = (\bar{1}00,100,\bar{1}00,100),
\end{equation}
and the spin configuration of other sites is generated from this and the ordering
wavevector.

\subsection{Role of pseudo-dipolar interaction: case 2}

Here we consider the  parameter regime with ${D=0}$, $ {\Gamma_1 = 0}$, $ {\Gamma_2 \neq 0}$. 
For ${\Gamma_2<0}$, the ground state is the same as the case for ${D<0}$, 
which is the 4-in-4-out state. For $\Gamma_2>0$, the anisotropy does not lift the classical 
degeneracy of the nearest-neighbor Heisenberg model on pyrochlore lattice.

\subsection{With both Dzyaloshinskii-Moriya and pseudo-dipolar interactions}
\label{sec:sec2D}

\begin{figure}[b]
\includegraphics[width=7.5cm]{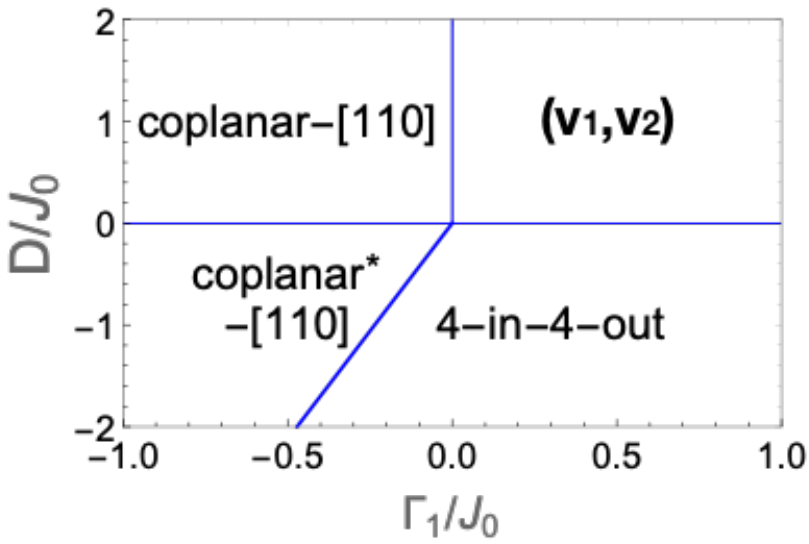}
\caption{The mean-field phase diagram in the $D$-$\Gamma_1$ plane with ${\Gamma_2= 0}$. 
The corresponding  spin configurations are found in Fig.~\ref{fig2}.}
\label{fig3}
\end{figure}

In this subsection, we study the classical phase diagram 
when both two of the anisotropic exchanges are present. 
We start from the $D$-$\Gamma_1$ plane with ${\Gamma_2 = 0}$.    
The phase diagram is depicted in Fig.~\ref{fig3}. In all the parts 
of the phase diagram, the magnetic wavevector is ${{\bf q} = {\bf 0}}$. 
Most parts of the phase diagram can be understood as the intersection 
of two different ground state manifolds separately favored by $D$ and 
$\Gamma_1$, which have already been discussed in detail in the previous subsections.

For ${D<0, \Gamma_1 >0}$, the 4-in-4-out state is favored. 
For ${D>0, \Gamma_1 >0}$,  we have the classical ground 
states constructed as the linear superpositions of the same two basis vectors ${\bf v}_1$ and ${\bf v}_2$ 
that are introduced in Eq.~\eqref{eq:basis1_dm} and ~\eqref{eq:basis2_dm} for the case of ${D>0}$. 
For ${D>0, \Gamma_1<0}$, the ground state is a coplanar state with the spins pointing along different [110] 
directions (denoted as ``coplanar-[110]'' in Fig.~\ref{fig3}), 
whose 4-spin vectors $\Psi$ can be constructed from the basis 
vectors ${\bf u}_1$ and ${\bf u}_2$ in Eq.~\eqref{eq:basiscp1} 
and Eq.~\eqref{eq:basiscp2},
\begin{equation}
\Psi = \frac{1}{\sqrt{2}} (110,\bar{1}10,1\bar{1}0,\bar{1}\bar{1}0).
\end{equation}

For ${D<0}, {\Gamma_1<0}$, the $D$-demanded and $\Gamma_1$-demanded 
ground state manifolds have no overlap. We find that, when ${D<3\sqrt{2}\Gamma_1}$,
DM interaction has a more weight in the Hamiltonian and the ground state is the 4-in-4-out 
state, and in the opposite case, the ground state is a coplanar state 
(denoted as ``coplanar$^{\ast}$-[110]'' in Fig.~\ref{fig3})
whose 4-spin vector is given
\begin{equation}
\Psi = \frac{1}{\sqrt{2}} (1\bar{1}0,\bar{1}\bar{1}0,110,\bar{1}10).
\label{eq:noncoplanar2}
\end{equation}
Note this coplanar state is distinct from the ``coplanar-[110]'' state found for $D>0, \Gamma_1<0$.

\begin{figure}[t]
\includegraphics[width=7.5cm]{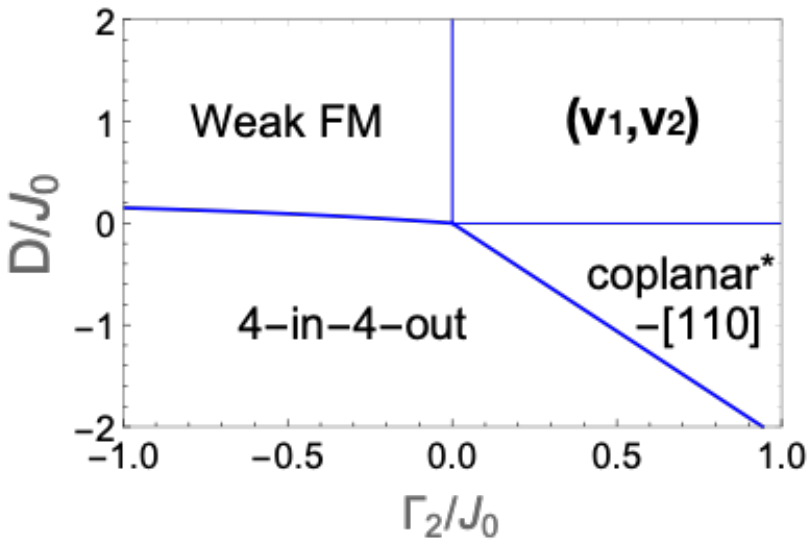}
\caption{ The mean-field phase diagram in the $D$-$\Gamma_2$ 
plane with ${\Gamma_1= 0}$. The corresponding 
spin configurations are found in Fig.~\ref{fig2}. }
\label{fig4}
\end{figure}

Now we discuss the ground states in $D$-$\Gamma_2$ plane with ${\Gamma_1 = 0}$. 
The phase diagram is depicted in Fig.~\ref{fig4}. The magnetic wavevector is 
${{\bf q} =0}$ everywhere in the phase diagram.

For ${D<0, \Gamma_2 <0}$, the ground state is simply the 4-in-4-out state. 
For ${D>0, \Gamma_2 >0}$, the ground state is an arbitrary linear superposition 
of the basis vectors ${\bf v}_1$ and ${\bf v}_2$ in Eq.~\eqref{eq:basis1_dm} 
and Eq.~\eqref{eq:basis2_dm}. In the regime of ${D>0, \Gamma_2<0}$, there 
exist two phases. When ${D>D_{c1}(\Gamma_2)}$ with 
\begin{equation}
D_{c1}(\Gamma_2) = \frac{\sqrt{2}}{6} (3J_0-2\Gamma_2 - \sqrt{9J_0^2 - 6J_0 \Gamma_2 +4 \Gamma_2^2}) ,
\end{equation}
the ground state turns out to be weakly ferromagnetic and denoted as 
``weak FM'' in Fig.~\ref{fig4}.  The 4-spin vectors of the magnetic unit cell are
parameterized as 
\begin{equation}
\Psi  = \cos \theta \, y_1 + \sin \theta \, y_2
\end{equation}
with 
\begin{eqnarray}
y_1 & = & \frac{1}{\sqrt{2}} (\bar{1}\bar{1}0,1\bar{1}0,\bar{1}10,110) ,
\\
y_2 & = & (001,001,001,001),
\end{eqnarray}
and the angular variable $\theta$ satisfies
\begin{eqnarray}
\cos 2 \theta &=& \frac{4J_0 +\sqrt{2}D -\Gamma_2 }{ \sqrt{(4J_0+\sqrt{2}D -\Gamma_2)^2 + 8\Gamma_2^2   }}
\\
\sin 2\theta   &=&  \frac{-2\sqrt{2} \Gamma_2 }{ \sqrt{(4J_0+\sqrt{2}D -\Gamma_2)^2 + 8 \Gamma_2^2   }}. 
\end{eqnarray}
When ${D< D_{c1} (\Gamma_2)}$, the ground state is the 4-in-4-out state.

In the region of ${D<0, \Gamma_2>0}$, there also exist two phases. 
When ${D < D_{c2} (\Gamma_2)}$ with $D_{c2} (\Gamma_2)$ given by
\begin{equation}
D_{c2} = -\frac{3\sqrt{2} \Gamma_2}{2}.
\end{equation}
When the DM interaction is dominant and negative, the ground state is the 4-in-4-out state. 
When ${D > D_{c2} (\Gamma_2)}$, a coplanar state with
spins pointing along various [110] directions is favored and the 4-spin 
vector $\Psi$ is the same as the one introduced in Eq.~\eqref{eq:noncoplanar2} 
and its symmetry equivalence. Hence, we also denote this coplanar state as 
``coplanar$^{\ast}$-[110]'' in Fig.~\ref{fig4}.

\section{Magnetism from electron quasi-itinerancy}
\label{sec3}

Having understood the role of each anisotropic exchange for the generic exchange Hamiltonian 
in the previous section, in this section we discuss the physical exchange Hamiltonian derived 
perturbatively from the microscopic parent Hubbard model and from there approach the 
magnetic states in the intermediate coupling regime. We analyze the possible magnetic 
ground states for the compound R$_2$Ir$_2$O$_7$.

\begin{figure}[b]
\includegraphics[width=8.6cm]{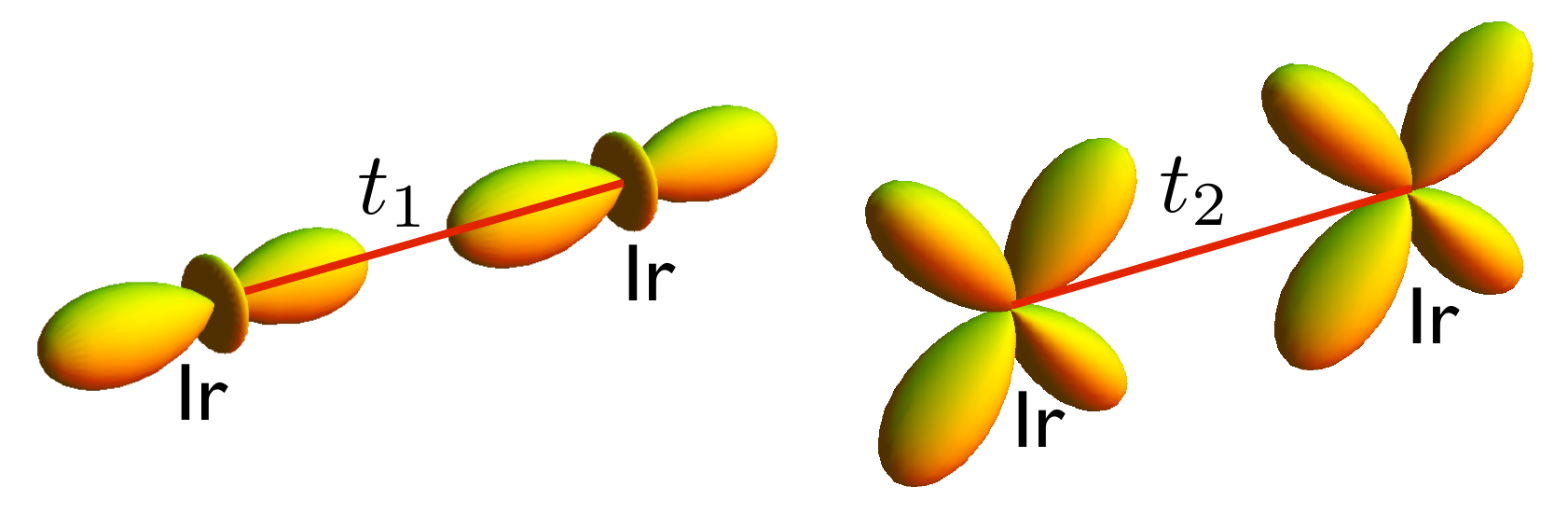}
\caption{(Color online.)  The direct electron tunneling between the Ir atoms. 
Left: the $\sigma$-bonding with tunneling amplitude $t_1$. 
Right: the $\pi$-bonding with tunneling amplitude $t_2$.}
\label{fig:bonds}
\end{figure}

\subsection{Hubbard model and exchange}

We assume the on-site SOC is strong enough so that the lower ${J=3/2}$ bands 
are completely filled and the upper ${J=1/2}$ bands are half filled.  
This approximation misses the hybridization between the ${J=1/2}$ and the
${J=3/2}$ bands, and this process may lead to some interesting properties 
and will be addressed in later works. The electrons can tunnel from one 
Ir$^{4+}$ ion to neighboring Ir$^{4+}$ ions either directly or 
indirectly via the $p$ orbitals of the intermediate oxygen ions~\cite{PesinBalents,PhysRevB.85.045124}. 
Since $5d$ electron orbitals are spatially extended, therefore the direct tunneling 
of electrons might be equally important as the indirect tunneling. 
With electrons locally projected onto the ${J=1/2}$ basis, 
one can write down a minimal Hubbard model~\cite{PhysRevB.85.045124}
\begin{equation}
{\mathcal H} =  \sum_{\langle ij \rangle}  \big[( {\mathcal T}^d_{ij,\alpha \beta}  
                     +  {\mathcal T}^{id}_{ij,\alpha \beta} ) d^{\dagger}_{i \alpha} d^{}_{j \beta} +h.c.\big]
                     +  \sum_i   U  n_{i, \uparrow}   n_{i,\downarrow} ,    
\label{eq:hubbard}
\end{equation}
in which, only the nearest-neighbor tunneling term is included,  
$d^{\dagger}_{i \alpha}$ ($d_{i\alpha}^{}$) is the creation (annihilation) operator  
for an electron on effective spin state $|{J=1/2, J^z =\alpha} \rangle$ at site $i$,
and ${n_{i\sigma} \equiv d^\dagger_{i\sigma} d^{}_{i\sigma}}$ measures the 
electron number with the spin $\sigma$ at site $i$. 
In Eq.~\eqref{eq:hubbard}, $ {\mathcal T}^d$ and ${\mathcal T}^{id}$ are 
the tunneling matrices for the direct and indirect processes, respectively.

\begin{figure}[t]
\includegraphics[width=7.5cm]{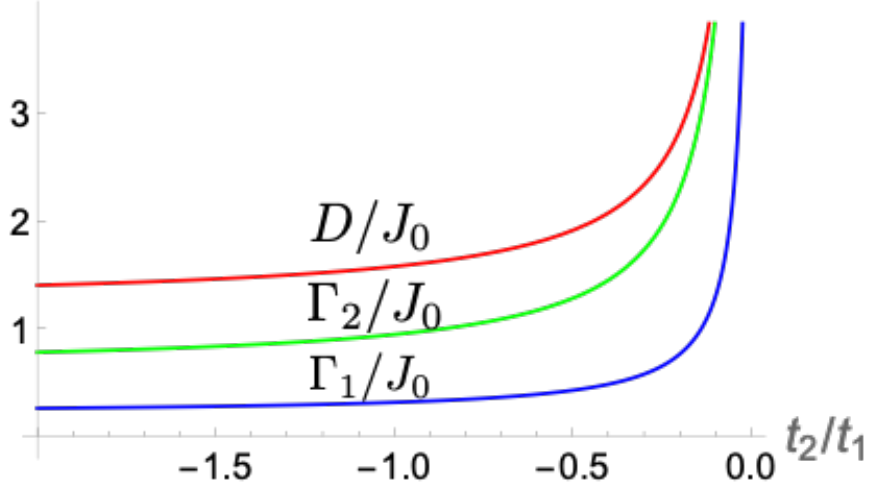}
\caption{(Color online.) The dependence of anisotropic couplings 
on the ratio between the $\pi$-bonding amplitude $t_2$ and the
$\sigma$-bonding amplitude $t_1$. In the plot, from top to bottom, 
the curves are $D/J_0$, $\Gamma_1/J_0$ and $\Gamma_2/J_0$. }
\label{fig6}
\end{figure}

For the direct tunneling processes, there exist two types of tunneling amplitudes: 
the $\sigma$-bonding $t_1$, and the $\pi$-bonding $t_2$ 
(see Fig.~\ref{fig:bonds})~\cite{PhysRevB.85.045124}.  Moreover, it is expected from the
orbital overlaps that $t_2$ has a different sign from $t_1$. 
In the limit of dominant direct tunneling, standard second order perturbation 
yields the exchange couplings introduced in Eq.~\eqref{eq:exchange},
\begin{eqnarray}
&&J_0  = \frac{603 t_1^2 -58296 t_1 t_2 + 248368 t_2^2}{2834352 U}
\\
&& D = \frac{ 5 \sqrt{2} (153 t_1^2 - 1356 t_1 t_2 + 2528 t_2^2)  }{118098 U}
\\
&&\Gamma_1 =  \frac{ 50 (9 t_1^2- 48  t_1t_2 + 64 t_2^2  )  }{ 177147U }
\\
&& \Gamma_2 =  3 \Gamma_1.
\label{eq:direct_ex}
\end{eqnarray}
It turns out that, the DM interaction has the most weight in the exchange Hamiltonian.
As $J_0$ is assured to be positive in Eq.~\eqref{eq:direct_ex}, we depict the ratios of  
$D/J_0, \Gamma_1/J_0$ and $\Gamma_2/J_0$ in Fig.~\ref{fig6}.

\begin{figure}[t]
\includegraphics[width=8cm]{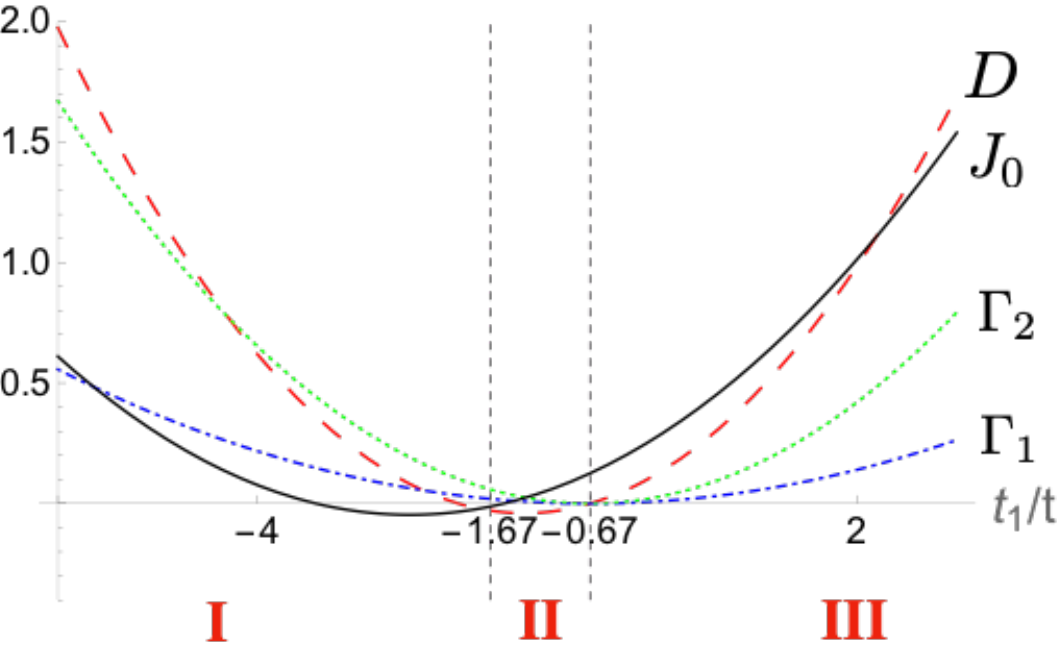}
\caption{(Color online.) The dependence of couplings on the ratio between the direct tunneling to the 
indirect tunneling. In the figure, we have set ${t_2 = -2t_1/3}$. The ground state of the region II is the 
``4-in-4-out'' state. For the region I and III, the ground state is degenerate and spanned by the basis 
vectors ${\bf v}_1$ and ${\bf v}_2$. The two dashed vertical lines are the phase boundaries separating 
the ``4-in-4-out'' state in region II from the $({\bf v}_1, {\bf v}_2)$ manifold in the region I and III. 
The unit of the vertical axis is set to be $t^2/U$.}
\label{fig7}
\end{figure}

In contrast, the indirect tunneling process is described by one single tunneling 
amplitude $t$.~\cite{PesinBalents} 
When it is dominant, the exchange couplings are given by
\begin{eqnarray}
&&J_0 =  \frac{49132 t^2 }{ 177147U } ,
\\
&& D  = \frac{ 7280 \sqrt{2} t^2 }{59049 U} ,
\\
&& \Gamma_1  =  \frac{1568 t^2  }{ 177147 U} ,
\\
&& \Gamma_2  =  3 \Gamma_1 .
\end{eqnarray}
It is important to note that, although we find ${\Gamma_2 =3 \Gamma_1}$ 
for both limits studied above, this relation is not protected by symmetry
 and will break down if a more realistic model is assumed. Although we find 
 that $J_0, D, \Gamma_1, \Gamma_2$ are all positive for the two limits studied above, 
this result actually breaks down when both direct and indirect tunnelings are included. 
As plotted in Fig.~\ref{fig7} for the case of ${t_2=-t_2/3}$, the Heisenberg exchange $J_0$ 
and DM interaction $D$ both change sign for certain intermediate ranges of $t_1/t$. 
This indicates that different magnetic order may emerge in the intermediate regimes of $t_1/t$.

\subsection{Ground states of the exchange Hamiltonian}
\label{sec3B}

In the previous subsection, we have explicitly derived the exchange Hamiltonian 
from the Hubbard model. For both exchanges in the limit of the dominant direct 
or indirect tunneling, the coupling parameters $J_0, D, \Gamma_1, \Gamma_2$   
are found to be positive. For this parameter regime, It is ready to show by
the mean-field theory and/or directly observe from the phase diagram
depicted in Fig.~\ref{fig3} and ~\ref{fig4} that, the mean-field classical ground state 
manifold is continuously degenerate and is spanned by the two basis vectors 
${\bf v}_1$ and ${\bf v}_2$ (see Eq.~\eqref{eq:basis1_dm} and Eq.~\eqref{eq:basis2_dm}). 
As shown in Fig.~\ref{fig7}, there is a region that the DM interaction $D$ 
changes sign, that may favor the ``4-in-4-out'' state as the classical ground state
in that region. After a complete calculation, we find the phase diagram that is 
depicted in Fig.~\ref{fig7}. Region II develops the ``4-in-4-out'' ground state. 
Region I and III have the degenerate ground state manifold $({\bf v}_1, {\bf v}_2)$. 
Remarkably, the phase boundary between region II and region I and III is exactly 
the same as the one obtained from a self-consistent mean field calculation
for the intermediate coupling regime in the calculation below and the
one in Ref.~\onlinecite{PhysRevB.85.045124}. 

\begin{figure}[t]
\includegraphics[width=8.4cm]{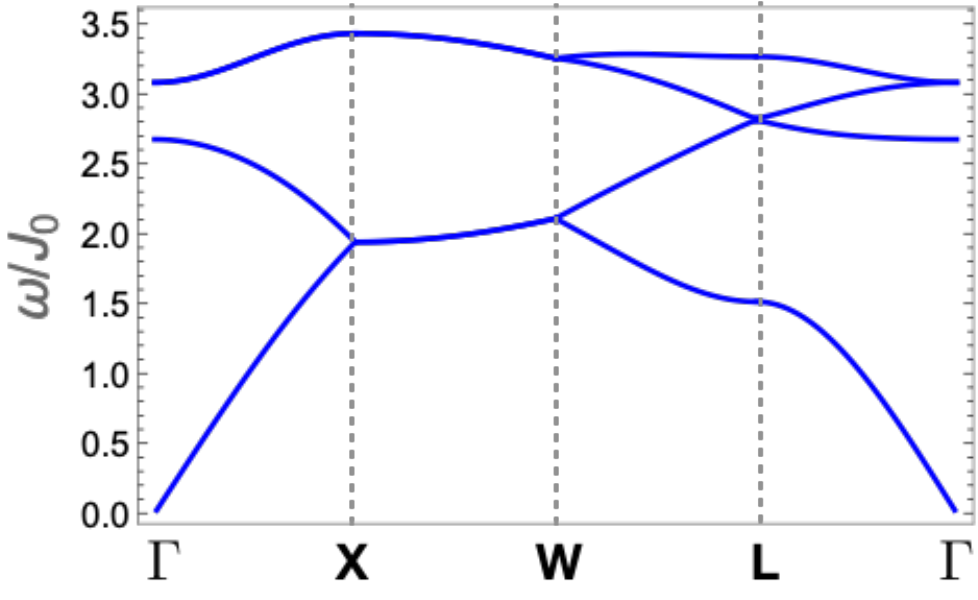}
\caption{The magnon dispersion along the high symmetry momentum direction 
$\Gamma$-X-W-L-$\Gamma$. 
The parameters in this figure are set to be ${D = 0.5 J_0}, {\Gamma_1 = 0.2 J_0}, {\Gamma_2 = 0.3 J_0}$. 
The gapless mode at the $\Gamma$ point is an artifact of the linear spin-wave theory. }
\label{fig8}
\end{figure}

This continuous degeneracy of $({\bf v}_1,{\bf v}_2)$ ground state manifold will be lifted 
if the quantum fluctuation is included.  We study this quantum order-by-disorder effect 
by the linear spin-wave theory. We express the classical 4-spin vectors as
\begin{equation}
\Psi = \cos \phi \, {\bf v}_1 + \sin \phi \, {\bf v}_2,
\end{equation} 
where $\phi$ parameterizes the orientation of the spin vectors. 
Then we introduce the Holstein-Primarkoff bosons,
\begin{eqnarray}
&& {\bf J}_i \cdot \hat{m}_i   =  J - a_i^{\dagger} a_i  , \\
&&{\bf J}_i \cdot \hat{n}_i    =  \frac{\sqrt{2J}}{2} ( a_i + a_i^{\dagger} ) ,    \\
&&{\bf J}_i \cdot (\hat{m}_i \times \hat{n}_i) =   \frac{\sqrt{2J}}{ 2 i} (a_i - a^{\dagger}_i),
\end{eqnarray}
here, $\hat{m}_i$ is the unit vector describing the spin orientation of classical spin order 
at site $i$, and $\hat{n}_i$ is a unit vector that is normal to $\hat{n}_i$ but within the
plane spanned by ${\bf v}_1$ and ${\bf v}_2$. Plugging the above relations into the exchange 
Hamiltonian, one is ready to write down the quadratic spin wave Hamiltonian,
\begin{eqnarray}
{\mathcal H}_{\text{sw}} &=&  \sum_{\bf k} \big[A_{ij} ({\bf k})  a_i^{\dagger} ({\bf k})a_j ({\bf k}) 
                                            + B_{ij} ({\bf k} ) a_i (-{\bf k})  a_j ({\bf k}) 
\nonumber \\
&&  \quad\quad   +   B^{\ast}_{ij} ({\bf k} ) a_i^{\dagger} ({\bf k})  a_j^{\dagger} (-{\bf k}) \big] + E_{\text{cl}} ,
\label{spinwave}
\end{eqnarray}
in which, $E_{\text{cl}}$ is the classical ground state energy, and $A_{ij}$ and $B_{ij}$ satisfy
\begin{eqnarray}
A_{ij} ({\bf k}) &=& A_{ij}^{\ast} ({\bf k}), \\
B_{ij} ({\bf k}) &=& B_{ji} (- {\bf k}),
\label{relation}
\end{eqnarray}
and are given in Appendix.~\ref{appendix}. From the quadratic spin-wave Hamiltonian, 
we obtain the quantum zero-point energy, that is found to be optimized by the non-coplanar 
spin configuration ${\bf v}_2$ (see Eq.~\eqref{eq:basis2_dm}) 
with ${\phi = \pi/2}$ (and its symmetry equivalences). 
We also find that, the magnon spectrum (see Fig.~\ref{fig8}) is gapless 
at the $\Gamma$ point, that originates from the continuous degeneracy of the classical 
ground states. This gapless mode is not supposed to remain if the anharmonic effects
beyond the linear spin-wave theory are included as the gapless feature is not protected
by any continuous symmetry of the Hamiltonian. A mini-gap would appear if a full 
calculation is performed. 

\subsection{Hubbard model and electron quasi-itinerancy 
in the intermediate coupling regime}
\label{sec3C} 

In the previous subsections, we have analyzed the magnetic ground states of the 
Ir-based pyrochlore lattice for R$_2$Ir$_2$O$_7$ in the strong coupling regime. 
We find that, even though the classical mean-field ground states are continuously 
degenerate for the exchange derived from the Hubbard model, all the ground 
states have a magnetic wavevector ${{\bf q} = {\bf 0}}$.   It is known that, the 
SOC twists the electron motion and reduces the electron bands. Although the 
large spatial extension of the $5d$ electrons reduces the electron correlation, 
as the bandwidth is also reduced, it is then not quite obvious where the actual 
physical system is located. Thus, it is legitimate for us to tackle the system 
from the strong correlation to the intermediate correlation by reducing the 
correlation strength. The knowledge that we have learned from the strong 
coupling regime may be extended to the intermediate regime.  
Moreover, the existing experiments on Eu$_2$Ir$_2$O$_7$, Nd$_2$Ir$_2$O$_7$,
Tb$_2$Ir$_2$O$_7$
and Sm$_2$Ir$_2$O$_7$ suggest a ${{\bf q} ={\bf 0}}$ magnetic order~\cite{PhysRevLett.117.037201,PhysRevB.89.140413,PhysRevMaterials.2.011402,PhysRevB.87.100403}. 
In this subsection, we study the magnetic properties of the Hubbard model in 
the intermediate coupling regime by a self-consistent mean field theory. Based on 
the results from the strong correlation regime, we assume the magnetic order 
in this regime also have a magnetic wavevector ${{\bf q} = {\bf 0}}$. 
To implement the mean-field theory, we decouple the Hubbard-$U$ interaction as,
\begin{eqnarray}
U n_{i, \uparrow} n_{i,\downarrow}  & = & -\frac{2U}{3}  {\bf J}_i^2  + \frac{U}{6} n_i
\nonumber \\
&\rightarrow & - \frac{2U}{3}  ( 2 \langle  {\bf J}_i  \rangle \cdot {\bf J}_i
                         - \langle {\bf J}_i \rangle^2 )  + \frac{U}{6} n_i,
\end{eqnarray}
in which, $n_i$ is the electron number at site $i$ and ${{\bf J}_i =  
\sum_{\alpha\beta}d^{\dagger}_{i\alpha} \boldsymbol{\sigma}^{}_{\alpha\beta} d^{}_{i\beta}/2}$ 
is the operator for the effective spin moment. With this decoupling, 
the mean-field Hamiltonian is quadratic with 
\begin{eqnarray}
H_{\text{MF}} &  \equiv  &
\sum_{\langle ij \rangle}  \big[( {\mathcal T}^d_{ij,\alpha \beta}  
                     +  {\mathcal T}^{id}_{ij,\alpha \beta} ) d^{\dagger}_{i \alpha} d^{}_{j \beta} +h.c.\big]
                     \nonumber \\
                   &&   -\sum_i  \frac{4U}{3}   \langle  {\bf J}_i  \rangle \cdot {\bf J}_i + \cdots,
\end{eqnarray}
where ``$\cdots$'' refers to the unessential terms that do not involve the electron operators.  
We then diagonalize the mean-field Hamiltonian and solve for the magnetic order of each 
sublattice self-consistently. Our results of the magnetic orders can be found in Fig.~\ref{fig2}. 
In the region II, the calculation quickly converges to the 4-in 4-out magnetic order. 
For the region I and the region III, the calculation does not quickly converge. After a few 
steps, the magnetic order from the self-consistent calculation actually drops into the 
continuous manifold that is spanned by the 4-spin vectors ${\bf v}_1$ and ${\bf v}_2$
and then fluctuate within this manifold without seeing a quick convergence. To resolve 
the magnetic orders in these two regions, we perform a different calculation below that 
may be illuminating. The self-consistent calculation tells us that the magnetic orders 
can be parameterized as 
\begin{eqnarray}
\Psi (\phi)&=& (  {\bf J}_0, {\bf J}_1, {\bf J}_2, {\bf J}_3) \nonumber \\
& = & M ( \cos \phi  \,{\bf v}_1 + \sin \phi \, {\bf v}_2 ),
\label{eq:mag}
\end{eqnarray}
where the order parameter $M$ depends on the dimensionless parameter $U/t$ 
that measures the strength of the interaction. For a given $U/t$, the magnetic 
order parameter $M$ is fixed. The self-consistent calculation was unable to quickly
converge the angular parameter $\phi$ which is the task to be fulfilled. It is ready to 
see that the task boils down to optimize the kinetic energy in the mean-field 
Hamiltonian $H_{\text{MF}}$, {\sl i.e.}
\begin{eqnarray}
\langle {\Psi ( \phi )} | 
\sum_{\langle ij \rangle}  \big[( {\mathcal T}^d_{ij,\alpha \beta}  
                     +  {\mathcal T}^{id}_{ij,\alpha \beta} ) d^{\dagger}_{i \alpha} d_{j \beta}^{} +h.c.\big]
|{\Psi ( \phi )}  \rangle .
\end{eqnarray}
The spirit of this calculation scheme is a bit similar to the double exchange. Over there, 
the itinerant electron is coupled with the local moments with ferromagnetic Kondo/Hund's coupling, 
and the magnetic order is established by optimizing the kinetic energy of the itinerant electrons 
and the exchange energy of the local moments~\cite{PhysRev.82.403}. 
In the doped maganites, to gain the kinetic energy, 
the local moments twist themselves from the spin configuration favored by the exchange energy. 
Another possibly electron kinetic energy driven magnetism was proposed for the doped 
van der Waals antiferromagnet CeTe$_3$~\cite{okuma2020fermionic}, and was refereed 
as fermionic order by disorder. 
For our case here, the electron kinetic energy is optimized within the background of 
the magnetism that operates on the continuously degenerate manifold. Our calculation 
suggests the selection of the angle $\phi$ to $\pi/2$ for all ${U>0}$. 
We find that, the kinetic energy stabilizes the non-coplanar state with
${\phi =\pi/2}$. Although this mechanism of breaking the continuous 
degeneracy by optimizing the kinetic energy is qualitatively different 
from the quantum order by disorder discussed in the previous subsection, 
the magnetic order from both mechanisms turns out to be identical, 
the phase boundaries separating different ordered phases are 
also remarkably identical for both mechanisms.  These results 
suggest that the magnetic orders in the intermediate and the strong 
coupling regimes may be continuously connected.


\section{Discussion}
\label{sec4}

To summarize, we have studied the magnetic ground states for the Ir-based 
pyrochlore lattice in both intermediate and strong coupling regimes. 
Various classical ground states are identified for the generic exchange 
Hamiltonian in the strong coupling limit. These results can be further 
applied to other magnetic systems on the pyrochlore lattice. We find that, 
the magnetic orders in the intermediate and strong coupling regimes for the 
pyrochlore iridates turn out to be identical.

The experiments on the pyrochlore iridates have rapidly evolved~\cite{PhysRevB.85.245109,PhysRevLett.109.136402,PhysRevB.85.214434,PhysRevB.86.014428,PhysRevB.85.205104,PhysRevB.87.100403,PhysRevLett.117.037201,PhysRevLett.115.056402,PhysRevB.89.140413,PhysRevB.89.115111,PhysRevLett.117.056403,PhysRevB.88.060411,PhysRevB.89.075127,PhysRevLett.114.247202,PhysRevB.87.060403,PhysRevB.90.235110,PhysRevB.83.180402,PhysRevB.94.161102,PhysRevB.92.094405,PhysRevB.96.094437,PhysRevMaterials.2.011402,2020arXiv200512768W,PhysRevB.101.121101,PhysRevB.96.144415}. There 
exists a large body of experimental works, and the review papers on
this topic can be more useful to the interested readers~\cite{2014ARCMP...5...57W,2016ARCMP...7..195R,2019arXiv190308081T,Schaffer_2016}. Instead of delving 
on a few specific experimental results and details, we here make some 
experimental suggestion based on the theoretical calculations in our work. 
In the strong coupling analysis, there exists a broad parameter regime
that the magnetic order is realized from the quantum order by disorder mechanism. 
Once the particular magnetic order with the ordering wavevector ${{\bf q} =0}$
and the spins orientating along the vector ${\bf v}_2$ in Eq.~\eqref{eq:basis2_dm}   
is realized, one can check if the excitation spectrum and thermodynamic properties
are consistent with the theoretical results. A qualitative feature in the magnetic 
excitation spectrum is the almost gapless mode at the $\Gamma$ point 
(see Fig.~\ref{fig8} and the explanation in Sec.~\ref{sec3C}). A consequence
on the thermodynamics is the a nearly $\sim T^3$ temperature dependence 
in the specific heat at the temperatures above the mini-gap energy. 
In the intermediate coupling scenario, the interaction and the charge gap are not very large 
compared to the bandwidth. Although the same magnetic order persists 
to the intermediate coupling regime, the quantum order by disorder mechanism 
are expected to break down. If one uses the local moment language and 
relies on the exchange interaction, one necessarily needs to invoke further neighbor 
exchanges and even the ring exchange interactions. These extra interactions modify
the original pairwise nearest-neighbor exchange model and will break the original 
applicability of the quantum order by disorder here. A surprising result in 
our self-consistent calculation in Sec.~\ref{sec3C} is that the magnetic order quickly 
falls into the degenerate manifold spanned by ${\bf v}_1$ and ${\bf v}_2$, and then we 
use the electron kinetic energy to break the degeneracy and select the magnetic order. 
This indicates that the degenerate manifold could be readily accessible 
if the system is activated by a small energy. A pump-probe measurement of the 
magnetic properties of the system would be helpful in this regards. 

Finally, the weak Mottness with quasi-itinerant electrons might be relevant for many 
other $4d$/$5d$ materials. The effect should be considered if the charge gap is not 
very large. It is very likely that many $4d$/$5d$ magnets would be located in this regime. 
Even the square lattice material Sr$_2$IrO$_4$ was believed to be proximate to a Mott 
transition~\cite{PhysRevB.57.R11039}.
The well-known $\alpha$-RuCl$_3$ has a relatively weak charge 
gap~\cite{PhysRevB.94.161106,PhysRevB.90.041112,PhysRevB.93.075144,PhysRevLett.117.126403}, 
even though the existing theoretical analysis mostly starts from a pairwise superexchange 
interaction between the effective spin-1/2 moments. The interlayer ring exchange, 
due to the weak Mott gap and the electron quasi-itinerancy, could be responsible 
for the anomalous thermal Hall effect in $\alpha$-RuCl$_3$ for the magnetic field 
in the honeycomb plane and parallel to the zig-zag ordering 
axis~\cite{Ong2020,2020arXiv200101899Y,PhysRevLett.120.217205}
where the interlayer magnetic flux could be experienced by the material.  
 


\acknowledgments

We thank Xu Ping Yao and Dr. Fei Ye Li for the help with Fig.~1 and Fig.~2. 
We acknowledge the hospitality of Aspen center for theoretical physics 
for the ultracold atom program in the summer of 2011 when and where this work was carried out. 
We especially thank Michael Hermele for discussion around that time period. 
GC is supported by the Ministry of Science and Technology of China 
with Grant No.2018YFGH000095, 2016YFA0301001, 2016YFA0300500, 
by Shanghai Municipal Science and Technology Major Project with Grant 
No.2019SHZDZX04, and by the Research Grants Council of Hong Kong 
with General Research Fund Grant No.17303819. The part of work in Boulder 
was supported by DOE award No.~desc0003910. 
XQW is supported by MOST 2016YFA0300501 and NSFC 11974244 
and additionally from a Shanghai talent program.

\appendix

\section{The linear spin-wave theory}
\label{appendix}

In Sec.~\ref{sec3B} of the main text, the couplings $A_{ij} ({\bf k})$ and $B_{ij} ({\bf k})$ in 
the spin wave Hamiltonian for the magnetic orders given by the basis vector ${\bf v}_2$ are 
listed as follows, 
\begin{eqnarray}
A_{00} ({\bf k}) &=& A_{11} ({\bf k})  \nonumber \\
&=& A_{22} ({\bf k} ) = A_{33} ({\bf k}) = c_1 ,
\\
A_{12} ({\bf k}) &=&  \frac{1}{24}(1+ e^{-i (k_y + k_z)/2 }) c_2, 
\\
A_{13} ({\bf k}) &=& \frac{1}{24} (1 + e^{ -i (k_x + k_z) /2} ) c_2,
\\
A_{14} ({\bf k}) &=& \frac{1}{12} (1 + e^{-i (k_x + k_y)/2}  ) c_3 ,
\\
A_{23} ({\bf k}) &=& \frac{1}{12} ( 1 + e^{-i (k_x - k_y)/2} ) c_3 ,
\\
A_{24} ({\bf k}) &=& \frac{1}{24} ( 1 + e^{-i (k_x - k_z)/2} ) c_2 ,
\\
A_{34} ({\bf k}) &=& \frac{1}{24} (1 + e^{- i (k_y - k_z) /2} ) c_2,
\end{eqnarray}
and
\begin{eqnarray}
B_{12} ({\bf k}) &=&  \frac{1}{24} (1+ e^{-i (k_y + k_z)/2 }) c_4      ,
\\
B_{13} ({\bf k}) &=& \frac{1}{24} (1+ e^{-i (k_x + k_z)/2 }) c_4^{\ast}  ,
\\
B_{14} ({\bf k}) &=& \frac{1}{12} (1+ e^{-i (k_x + k_y)/2 }) c_5 ,
\\
B_{23} ({\bf k}) &=& \frac{1}{12} ( 1 + e^{- i (k_x - k_y)/2}  ) c_5 , 
\\
B_{24} ( {\bf k}) &=& \frac{1}{24} (1+ e^{-i (k_x - k_z)/2 }) c_4^{\ast}, 
\\
B_{34} ({\bf k}) &=& \frac{1}{24} (1+ e^{-i (k_y - k_z)/2 }) c_4,
\end{eqnarray}
in which, we have set ${J=1/2}$ and the coefficients are given as
\begin{eqnarray}
c_1 &=& J_0 +\sqrt{2} D + 4 \Gamma_1 + \Gamma_2,
 \\
c_2 &=& -2 J_0 + \sqrt{2} D - 17\Gamma_1 + 4 \Gamma_2,
 \\
c_3 &=& -4 J_0  +2 \sqrt{2} D - 7\Gamma_1 -\Gamma_2 ,
\\
c_4 &=& - (2 + 4i \sqrt{6} ) J_0 + (7\sqrt{2} - 2i\sqrt{3}) D \nonumber \\
        &  & +  (1+ 2i\sqrt{6} ) \Gamma_1 + (4+ 2i \sqrt{6} ) \Gamma_2 ,
\\
c_5 &=&   2J_0 + 2\sqrt{2} D - \Gamma_1 + 5 \Gamma_2 .
\end{eqnarray}
Other entries of $A_{ij} ( {\bf k} )$ and $B_{ij} ({\bf k})$ are either zero 
or obtained by the relations in Eq.~\eqref{relation}.
 
\bibliography{ref}

\begin{thebibliography}{72}%
\makeatletter
\providecommand \@ifxundefined [1]{%
 \@ifx{#1\undefined}
}%
\providecommand \@ifnum [1]{%
 \ifnum #1\expandafter \@firstoftwo
 \else \expandafter \@secondoftwo
 \fi
}%
\providecommand \@ifx [1]{%
 \ifx #1\expandafter \@firstoftwo
 \else \expandafter \@secondoftwo
 \fi
}%
\providecommand \natexlab [1]{#1}%
\providecommand \enquote  [1]{``#1''}%
\providecommand \bibnamefont  [1]{#1}%
\providecommand \bibfnamefont [1]{#1}%
\providecommand \citenamefont [1]{#1}%
\providecommand \href@noop [0]{\@secondoftwo}%
\providecommand \href [0]{\begingroup \@sanitize@url \@href}%
\providecommand \@href[1]{\@@startlink{#1}\@@href}%
\providecommand \@@href[1]{\endgroup#1\@@endlink}%
\providecommand \@sanitize@url [0]{\catcode `\\12\catcode `\$12\catcode
  `\&12\catcode `\#12\catcode `\^12\catcode `\_12\catcode `\%12\relax}%
\providecommand \@@startlink[1]{}%
\providecommand \@@endlink[0]{}%
\providecommand \url  [0]{\begingroup\@sanitize@url \@url }%
\providecommand \@url [1]{\endgroup\@href {#1}{\urlprefix }}%
\providecommand \urlprefix  [0]{URL }%
\providecommand \Eprint [0]{\href }%
\providecommand \doibase [0]{http://dx.doi.org/}%
\providecommand \selectlanguage [0]{\@gobble}%
\providecommand \bibinfo  [0]{\@secondoftwo}%
\providecommand \bibfield  [0]{\@secondoftwo}%
\providecommand \translation [1]{[#1]}%
\providecommand \BibitemOpen [0]{}%
\providecommand \bibitemStop [0]{}%
\providecommand \bibitemNoStop [0]{.\EOS\space}%
\providecommand \EOS [0]{\spacefactor3000\relax}%
\providecommand \BibitemShut  [1]{\csname bibitem#1\endcsname}%
\let\auto@bib@innerbib\@empty
\bibitem [{\citenamefont {{Witczak-Krempa}}\ \emph {et~al.}(2014)\citenamefont
  {{Witczak-Krempa}}, \citenamefont {{Chen}}, \citenamefont {{Kim}},\ and\
  \citenamefont {{Balents}}}]{2014ARCMP...5...57W}%
  \BibitemOpen
  \bibfield  {author} {\bibinfo {author} {\bibfnamefont {William}\ \bibnamefont
  {{Witczak-Krempa}}}, \bibinfo {author} {\bibfnamefont {Gang}\ \bibnamefont
  {{Chen}}}, \bibinfo {author} {\bibfnamefont {Yong~Baek}\ \bibnamefont
  {{Kim}}}, \ and\ \bibinfo {author} {\bibfnamefont {Leon}\ \bibnamefont
  {{Balents}}},\ }\bibfield  {title} {\enquote {\bibinfo {title} {{Correlated
  Quantum Phenomena in the Strong Spin-Orbit Regime}},}\ }\href {\doibase
  10.1146/annurev-conmatphys-020911-125138} {\bibfield  {journal} {\bibinfo
  {journal} {Annual Review of Condensed Matter Physics}\ }\textbf {\bibinfo
  {volume} {5}},\ \bibinfo {pages} {57--82} (\bibinfo {year} {2014})},\ \Eprint
  {http://arxiv.org/abs/1305.2193} {arXiv:1305.2193 [cond-mat.str-el]}
  \BibitemShut {NoStop}%
\bibitem [{\citenamefont {{Rau}}\ \emph {et~al.}(2016)\citenamefont {{Rau}},
  \citenamefont {{Lee}},\ and\ \citenamefont {{Kee}}}]{2016ARCMP...7..195R}%
  \BibitemOpen
  \bibfield  {author} {\bibinfo {author} {\bibfnamefont {Jeffrey~G.}\
  \bibnamefont {{Rau}}}, \bibinfo {author} {\bibfnamefont {Eric Kin-Ho}\
  \bibnamefont {{Lee}}}, \ and\ \bibinfo {author} {\bibfnamefont {Hae-Young}\
  \bibnamefont {{Kee}}},\ }\bibfield  {title} {\enquote {\bibinfo {title}
  {{Spin-Orbit Physics Giving Rise to Novel Phases in Correlated Systems:
  Iridates and Related Materials}},}\ }\href {\doibase
  10.1146/annurev-conmatphys-031115-011319} {\bibfield  {journal} {\bibinfo
  {journal} {Annual Review of Condensed Matter Physics}\ }\textbf {\bibinfo
  {volume} {7}},\ \bibinfo {pages} {195--221} (\bibinfo {year} {2016})},\
  \Eprint {http://arxiv.org/abs/1507.06323} {arXiv:1507.06323
  [cond-mat.str-el]} \BibitemShut {NoStop}%
\bibitem [{\citenamefont {{Takagi}}\ \emph {et~al.}(2019)\citenamefont
  {{Takagi}}, \citenamefont {{Takayama}}, \citenamefont {{Jackeli}},
  \citenamefont {{Khaliullin}},\ and\ \citenamefont
  {{Nagler}}}]{2019arXiv190308081T}%
  \BibitemOpen
  \bibfield  {author} {\bibinfo {author} {\bibfnamefont {Hidenori}\
  \bibnamefont {{Takagi}}}, \bibinfo {author} {\bibfnamefont {Tomohiro}\
  \bibnamefont {{Takayama}}}, \bibinfo {author} {\bibfnamefont {George}\
  \bibnamefont {{Jackeli}}}, \bibinfo {author} {\bibfnamefont {Giniyat}\
  \bibnamefont {{Khaliullin}}}, \ and\ \bibinfo {author} {\bibfnamefont
  {Stephen~E.}\ \bibnamefont {{Nagler}}},\ }\bibfield  {title} {\enquote
  {\bibinfo {title} {{Concept and realization of Kitaev quantum spin
  liquids}},}\ }\href@noop {} {\bibfield  {journal} {\bibinfo  {journal}
  {Nature Reviews Physics}\ }\textbf {\bibinfo {volume} {1}},\ \bibinfo {pages}
  {264–280} (\bibinfo {year} {2019})}\BibitemShut {NoStop}%
\bibitem [{\citenamefont {Schaffer}\ \emph {et~al.}(2016)\citenamefont
  {Schaffer}, \citenamefont {Lee}, \citenamefont {Yang},\ and\ \citenamefont
  {Kim}}]{Schaffer_2016}%
  \BibitemOpen
  \bibfield  {author} {\bibinfo {author} {\bibfnamefont {Robert}\ \bibnamefont
  {Schaffer}}, \bibinfo {author} {\bibfnamefont {Eric Kin-Ho}\ \bibnamefont
  {Lee}}, \bibinfo {author} {\bibfnamefont {Bohm-Jung}\ \bibnamefont {Yang}}, \
  and\ \bibinfo {author} {\bibfnamefont {Yong~Baek}\ \bibnamefont {Kim}},\
  }\bibfield  {title} {\enquote {\bibinfo {title} {{Recent progress on
  correlated electron systems with strong spin{\textendash}orbit coupling}},}\
  }\href {\doibase 10.1088/0034-4885/79/9/094504} {\bibfield  {journal}
  {\bibinfo  {journal} {Reports on Progress in Physics}\ }\textbf {\bibinfo
  {volume} {79}},\ \bibinfo {pages} {094504} (\bibinfo {year}
  {2016})}\BibitemShut {NoStop}%
\bibitem [{\citenamefont {Okamoto}\ \emph {et~al.}(2007)\citenamefont
  {Okamoto}, \citenamefont {Nohara}, \citenamefont {Aruga-Katori},\ and\
  \citenamefont {Takagi}}]{PhysRevLett.99.137207}%
  \BibitemOpen
  \bibfield  {author} {\bibinfo {author} {\bibfnamefont {Yoshihiko}\
  \bibnamefont {Okamoto}}, \bibinfo {author} {\bibfnamefont {Minoru}\
  \bibnamefont {Nohara}}, \bibinfo {author} {\bibfnamefont {Hiroko}\
  \bibnamefont {Aruga-Katori}}, \ and\ \bibinfo {author} {\bibfnamefont
  {Hidenori}\ \bibnamefont {Takagi}},\ }\bibfield  {title} {\enquote {\bibinfo
  {title} {{Spin-Liquid State in the $S=1/2$ Hyperkagome Antiferromagnet
  ${\mathrm{Na}}_{4}{\mathrm{Ir}}_{3}{\mathrm{O}}_{8}$}},}\ }\href {\doibase
  10.1103/PhysRevLett.99.137207} {\bibfield  {journal} {\bibinfo  {journal}
  {Phys. Rev. Lett.}\ }\textbf {\bibinfo {volume} {99}},\ \bibinfo {pages}
  {137207} (\bibinfo {year} {2007})}\BibitemShut {NoStop}%
\bibitem [{\citenamefont {Cao}\ \emph {et~al.}(1998)\citenamefont {Cao},
  \citenamefont {Bolivar}, \citenamefont {McCall}, \citenamefont {Crow},\ and\
  \citenamefont {Guertin}}]{PhysRevB.57.R11039}%
  \BibitemOpen
  \bibfield  {author} {\bibinfo {author} {\bibfnamefont {G.}~\bibnamefont
  {Cao}}, \bibinfo {author} {\bibfnamefont {J.}~\bibnamefont {Bolivar}},
  \bibinfo {author} {\bibfnamefont {S.}~\bibnamefont {McCall}}, \bibinfo
  {author} {\bibfnamefont {J.~E.}\ \bibnamefont {Crow}}, \ and\ \bibinfo
  {author} {\bibfnamefont {R.~P.}\ \bibnamefont {Guertin}},\ }\bibfield
  {title} {\enquote {\bibinfo {title} {{Weak ferromagnetism, metal-to-nonmetal
  transition, and negative differential resistivity in single-crystal
  ${\mathrm{Sr}}_{2}{\mathrm{IrO}}_{4}$}},}\ }\href {\doibase
  10.1103/PhysRevB.57.R11039} {\bibfield  {journal} {\bibinfo  {journal} {Phys.
  Rev. B}\ }\textbf {\bibinfo {volume} {57}},\ \bibinfo {pages}
  {R11039--R11042} (\bibinfo {year} {1998})}\BibitemShut {NoStop}%
\bibitem [{\citenamefont {Chen}\ and\ \citenamefont
  {Balents}(2008)}]{hyperk_chen}%
  \BibitemOpen
  \bibfield  {author} {\bibinfo {author} {\bibfnamefont {Gang}\ \bibnamefont
  {Chen}}\ and\ \bibinfo {author} {\bibfnamefont {Leon}\ \bibnamefont
  {Balents}},\ }\bibfield  {title} {\enquote {\bibinfo {title} {Spin-orbit
  effects in ${\text{na}}_{4}{\text{ir}}_{3}{\text{o}}_{8}$: A hyper-kagome
  lattice antiferromagnet},}\ }\href {\doibase 10.1103/PhysRevB.78.094403}
  {\bibfield  {journal} {\bibinfo  {journal} {Phys. Rev. B}\ }\textbf {\bibinfo
  {volume} {78}},\ \bibinfo {pages} {094403} (\bibinfo {year}
  {2008})}\BibitemShut {NoStop}%
\bibitem [{\citenamefont {Jackeli}\ and\ \citenamefont
  {Khaliullin}(2009)}]{PhysRevLett.102.017205}%
  \BibitemOpen
  \bibfield  {author} {\bibinfo {author} {\bibfnamefont {G.}~\bibnamefont
  {Jackeli}}\ and\ \bibinfo {author} {\bibfnamefont {G.}~\bibnamefont
  {Khaliullin}},\ }\bibfield  {title} {\enquote {\bibinfo {title} {{Mott
  Insulators in the Strong Spin-Orbit Coupling Limit: From Heisenberg to a
  Quantum Compass and Kitaev Models}},}\ }\href {\doibase
  10.1103/PhysRevLett.102.017205} {\bibfield  {journal} {\bibinfo  {journal}
  {Phys. Rev. Lett.}\ }\textbf {\bibinfo {volume} {102}},\ \bibinfo {pages}
  {017205} (\bibinfo {year} {2009})}\BibitemShut {NoStop}%
\bibitem [{\citenamefont {Kim}\ \emph {et~al.}(2009)\citenamefont {Kim},
  \citenamefont {Ohsumi}, \citenamefont {Komesu}, \citenamefont {Sakai},
  \citenamefont {Morita}, \citenamefont {Takagi},\ and\ \citenamefont
  {Arima}}]{xray}%
  \BibitemOpen
  \bibfield  {author} {\bibinfo {author} {\bibfnamefont {B.~J.}\ \bibnamefont
  {Kim}}, \bibinfo {author} {\bibfnamefont {H.}~\bibnamefont {Ohsumi}},
  \bibinfo {author} {\bibfnamefont {T.}~\bibnamefont {Komesu}}, \bibinfo
  {author} {\bibfnamefont {S.}~\bibnamefont {Sakai}}, \bibinfo {author}
  {\bibfnamefont {T.}~\bibnamefont {Morita}}, \bibinfo {author} {\bibfnamefont
  {H.}~\bibnamefont {Takagi}}, \ and\ \bibinfo {author} {\bibfnamefont
  {T.}~\bibnamefont {Arima}},\ }\bibfield  {title} {\enquote {\bibinfo {title}
  {{Phase-Sensitive Observation of a Spin-Orbital Mott State in
  Sr$_2$IrO$_4$}},}\ }\href {\doibase 10.1126/science.1167106} {\bibfield
  {journal} {\bibinfo  {journal} {Science}\ }\textbf {\bibinfo {volume}
  {323}},\ \bibinfo {pages} {1329} (\bibinfo {year} {2009})}\BibitemShut
  {NoStop}%
\bibitem [{\citenamefont {Chaloupka}\ \emph {et~al.}(2010)\citenamefont
  {Chaloupka}, \citenamefont {Jackeli},\ and\ \citenamefont
  {Khaliullin}}]{PhysRevLett.105.027204}%
  \BibitemOpen
  \bibfield  {author} {\bibinfo {author} {\bibfnamefont {Ji\ifmmode
  \check{r}\else~\v{r}\fi{}\'{\i}}\ \bibnamefont {Chaloupka}}, \bibinfo
  {author} {\bibfnamefont {George}\ \bibnamefont {Jackeli}}, \ and\ \bibinfo
  {author} {\bibfnamefont {Giniyat}\ \bibnamefont {Khaliullin}},\ }\bibfield
  {title} {\enquote {\bibinfo {title} {{Kitaev-Heisenberg Model on a Honeycomb
  Lattice: Possible Exotic Phases in Iridium Oxides
  ${A}_{2}{\mathrm{IrO}}_{3}$}},}\ }\href {\doibase
  10.1103/PhysRevLett.105.027204} {\bibfield  {journal} {\bibinfo  {journal}
  {Phys. Rev. Lett.}\ }\textbf {\bibinfo {volume} {105}},\ \bibinfo {pages}
  {027204} (\bibinfo {year} {2010})}\BibitemShut {NoStop}%
\bibitem [{\citenamefont {Lee}\ and\ \citenamefont
  {Lee}(2005)}]{PhysRevLett.95.036403}%
  \BibitemOpen
  \bibfield  {author} {\bibinfo {author} {\bibfnamefont {Sung-Sik}\
  \bibnamefont {Lee}}\ and\ \bibinfo {author} {\bibfnamefont {Patrick~A.}\
  \bibnamefont {Lee}},\ }\bibfield  {title} {\enquote {\bibinfo {title} {{U(1)
  Gauge Theory of the Hubbard Model: Spin Liquid States and Possible
  Application to
  $\ensuremath{\kappa}\mathrm{\text{\ensuremath{-}}}(\mathrm{BEDT}\mathrm{\text{\ensuremath{-}}}\mathrm{TTF}{)}_{2}{\mathrm{Cu}}_{2}(\mathrm{CN}{)}_{3}$}},}\
  }\href {\doibase 10.1103/PhysRevLett.95.036403} {\bibfield  {journal}
  {\bibinfo  {journal} {Phys. Rev. Lett.}\ }\textbf {\bibinfo {volume} {95}},\
  \bibinfo {pages} {036403} (\bibinfo {year} {2005})}\BibitemShut {NoStop}%
\bibitem [{\citenamefont {Kato}\ \emph {et~al.}(2012)\citenamefont {Kato},
  \citenamefont {Chern}, \citenamefont {Al-Hassanieh}, \citenamefont
  {Perkins},\ and\ \citenamefont {Batista}}]{PhysRevLett.108.247215}%
  \BibitemOpen
  \bibfield  {author} {\bibinfo {author} {\bibfnamefont {Yasuyuki}\
  \bibnamefont {Kato}}, \bibinfo {author} {\bibfnamefont {Gia-Wei}\
  \bibnamefont {Chern}}, \bibinfo {author} {\bibfnamefont {K.~A.}\ \bibnamefont
  {Al-Hassanieh}}, \bibinfo {author} {\bibfnamefont {Natalia~B.}\ \bibnamefont
  {Perkins}}, \ and\ \bibinfo {author} {\bibfnamefont {C.~D.}\ \bibnamefont
  {Batista}},\ }\bibfield  {title} {\enquote {\bibinfo {title} {{Orbital
  Disorder Induced by Charge Fluctuations in Vanadium Spinels}},}\ }\href
  {\doibase 10.1103/PhysRevLett.108.247215} {\bibfield  {journal} {\bibinfo
  {journal} {Phys. Rev. Lett.}\ }\textbf {\bibinfo {volume} {108}},\ \bibinfo
  {pages} {247215} (\bibinfo {year} {2012})}\BibitemShut {NoStop}%
\bibitem [{\citenamefont {Chern}\ and\ \citenamefont
  {Batista}(2011)}]{PhysRevLett.107.186403}%
  \BibitemOpen
  \bibfield  {author} {\bibinfo {author} {\bibfnamefont {Gia-Wei}\ \bibnamefont
  {Chern}}\ and\ \bibinfo {author} {\bibfnamefont {Cristian~D.}\ \bibnamefont
  {Batista}},\ }\bibfield  {title} {\enquote {\bibinfo {title} {{Spin
  Superstructure and Noncoplanar Ordering in Metallic Pyrochlore Magnets with
  Degenerate Orbitals}},}\ }\href {\doibase 10.1103/PhysRevLett.107.186403}
  {\bibfield  {journal} {\bibinfo  {journal} {Phys. Rev. Lett.}\ }\textbf
  {\bibinfo {volume} {107}},\ \bibinfo {pages} {186403} (\bibinfo {year}
  {2011})}\BibitemShut {NoStop}%
\bibitem [{\citenamefont {Khomskii}\ and\ \citenamefont
  {Mizokawa}(2005)}]{PhysRevLett.94.156402}%
  \BibitemOpen
  \bibfield  {author} {\bibinfo {author} {\bibfnamefont {D.~I.}\ \bibnamefont
  {Khomskii}}\ and\ \bibinfo {author} {\bibfnamefont {T.}~\bibnamefont
  {Mizokawa}},\ }\bibfield  {title} {\enquote {\bibinfo {title} {{Orbitally
  Induced Peierls State in Spinels}},}\ }\href {\doibase
  10.1103/PhysRevLett.94.156402} {\bibfield  {journal} {\bibinfo  {journal}
  {Phys. Rev. Lett.}\ }\textbf {\bibinfo {volume} {94}},\ \bibinfo {pages}
  {156402} (\bibinfo {year} {2005})}\BibitemShut {NoStop}%
\bibitem [{\citenamefont {Pesin}\ and\ \citenamefont
  {Balents}(2010)}]{PesinBalents}%
  \BibitemOpen
  \bibfield  {author} {\bibinfo {author} {\bibfnamefont {D.}~\bibnamefont
  {Pesin}}\ and\ \bibinfo {author} {\bibfnamefont {L.}~\bibnamefont
  {Balents}},\ }\bibfield  {title} {\enquote {\bibinfo {title} {Mott physics
  and band topology in materials with strong spin-orbit interaction},}\ }\href
  {\doibase 10.1038/NPHYS1606} {\bibfield  {journal} {\bibinfo  {journal}
  {Nature Physics}\ }\textbf {\bibinfo {volume} {6}},\ \bibinfo {pages} {376}
  (\bibinfo {year} {2010})}\BibitemShut {NoStop}%
\bibitem [{\citenamefont {Go}\ \emph {et~al.}(2012)\citenamefont {Go},
  \citenamefont {Witczak-Krempa}, \citenamefont {Jeon}, \citenamefont {Park},\
  and\ \citenamefont {Kim}}]{PhysRevLett.109.066401}%
  \BibitemOpen
  \bibfield  {author} {\bibinfo {author} {\bibfnamefont {Ara}\ \bibnamefont
  {Go}}, \bibinfo {author} {\bibfnamefont {William}\ \bibnamefont
  {Witczak-Krempa}}, \bibinfo {author} {\bibfnamefont {Gun~Sang}\ \bibnamefont
  {Jeon}}, \bibinfo {author} {\bibfnamefont {Kwon}\ \bibnamefont {Park}}, \
  and\ \bibinfo {author} {\bibfnamefont {Yong~Baek}\ \bibnamefont {Kim}},\
  }\bibfield  {title} {\enquote {\bibinfo {title} {{Correlation Effects on 3D
  Topological Phases: From Bulk to Boundary}},}\ }\href {\doibase
  10.1103/PhysRevLett.109.066401} {\bibfield  {journal} {\bibinfo  {journal}
  {Phys. Rev. Lett.}\ }\textbf {\bibinfo {volume} {109}},\ \bibinfo {pages}
  {066401} (\bibinfo {year} {2012})}\BibitemShut {NoStop}%
\bibitem [{\citenamefont {Witczak-Krempa}\ and\ \citenamefont
  {Kim}(2012)}]{PhysRevB.85.045124}%
  \BibitemOpen
  \bibfield  {author} {\bibinfo {author} {\bibfnamefont {William}\ \bibnamefont
  {Witczak-Krempa}}\ and\ \bibinfo {author} {\bibfnamefont {Yong~Baek}\
  \bibnamefont {Kim}},\ }\bibfield  {title} {\enquote {\bibinfo {title}
  {{Topological and magnetic phases of interacting electrons in the pyrochlore
  iridates}},}\ }\href {\doibase 10.1103/PhysRevB.85.045124} {\bibfield
  {journal} {\bibinfo  {journal} {Phys. Rev. B}\ }\textbf {\bibinfo {volume}
  {85}},\ \bibinfo {pages} {045124} (\bibinfo {year} {2012})}\BibitemShut
  {NoStop}%
\bibitem [{\citenamefont {Wan}\ \emph {et~al.}(2011)\citenamefont {Wan},
  \citenamefont {Turner}, \citenamefont {Vishwanath},\ and\ \citenamefont
  {Savrasov}}]{PhysRevB.83.205101}%
  \BibitemOpen
  \bibfield  {author} {\bibinfo {author} {\bibfnamefont {Xiangang}\
  \bibnamefont {Wan}}, \bibinfo {author} {\bibfnamefont {Ari~M.}\ \bibnamefont
  {Turner}}, \bibinfo {author} {\bibfnamefont {Ashvin}\ \bibnamefont
  {Vishwanath}}, \ and\ \bibinfo {author} {\bibfnamefont {Sergey~Y.}\
  \bibnamefont {Savrasov}},\ }\bibfield  {title} {\enquote {\bibinfo {title}
  {Topological semimetal and fermi-arc surface states in the electronic
  structure of pyrochlore iridates},}\ }\href {\doibase
  10.1103/PhysRevB.83.205101} {\bibfield  {journal} {\bibinfo  {journal} {Phys.
  Rev. B}\ }\textbf {\bibinfo {volume} {83}},\ \bibinfo {pages} {205101}
  (\bibinfo {year} {2011})}\BibitemShut {NoStop}%
\bibitem [{\citenamefont {Ishikawa}\ \emph {et~al.}(2012)\citenamefont
  {Ishikawa}, \citenamefont {O'Farrell},\ and\ \citenamefont
  {Nakatsuji}}]{PhysRevB.85.245109}%
  \BibitemOpen
  \bibfield  {author} {\bibinfo {author} {\bibfnamefont {Jun~J.}\ \bibnamefont
  {Ishikawa}}, \bibinfo {author} {\bibfnamefont {Eoin C.~T.}\ \bibnamefont
  {O'Farrell}}, \ and\ \bibinfo {author} {\bibfnamefont {Satoru}\ \bibnamefont
  {Nakatsuji}},\ }\bibfield  {title} {\enquote {\bibinfo {title} {{Continuous
  transition between antiferromagnetic insulator and paramagnetic metal in the
  pyrochlore iridate Eu${}_{2}$Ir${}_{2}$O${}_{7}$}},}\ }\href {\doibase
  10.1103/PhysRevB.85.245109} {\bibfield  {journal} {\bibinfo  {journal} {Phys.
  Rev. B}\ }\textbf {\bibinfo {volume} {85}},\ \bibinfo {pages} {245109}
  (\bibinfo {year} {2012})}\BibitemShut {NoStop}%
\bibitem [{\citenamefont {Ueda}\ \emph {et~al.}(2012)\citenamefont {Ueda},
  \citenamefont {Fujioka}, \citenamefont {Takahashi}, \citenamefont {Suzuki},
  \citenamefont {Ishiwata}, \citenamefont {Taguchi},\ and\ \citenamefont
  {Tokura}}]{PhysRevLett.109.136402}%
  \BibitemOpen
  \bibfield  {author} {\bibinfo {author} {\bibfnamefont {K.}~\bibnamefont
  {Ueda}}, \bibinfo {author} {\bibfnamefont {J.}~\bibnamefont {Fujioka}},
  \bibinfo {author} {\bibfnamefont {Y.}~\bibnamefont {Takahashi}}, \bibinfo
  {author} {\bibfnamefont {T.}~\bibnamefont {Suzuki}}, \bibinfo {author}
  {\bibfnamefont {S.}~\bibnamefont {Ishiwata}}, \bibinfo {author}
  {\bibfnamefont {Y.}~\bibnamefont {Taguchi}}, \ and\ \bibinfo {author}
  {\bibfnamefont {Y.}~\bibnamefont {Tokura}},\ }\bibfield  {title} {\enquote
  {\bibinfo {title} {{Variation of Charge Dynamics in the Course of
  Metal-Insulator Transition for Pyrochlore-Type
  ${\mathrm{Nd}}_{2}{\mathrm{Ir}}_{2}{\mathbf{O}}_{7}$}},}\ }\href {\doibase
  10.1103/PhysRevLett.109.136402} {\bibfield  {journal} {\bibinfo  {journal}
  {Phys. Rev. Lett.}\ }\textbf {\bibinfo {volume} {109}},\ \bibinfo {pages}
  {136402} (\bibinfo {year} {2012})}\BibitemShut {NoStop}%
\bibitem [{\citenamefont {Shapiro}\ \emph {et~al.}(2012)\citenamefont
  {Shapiro}, \citenamefont {Riggs}, \citenamefont {Stone}, \citenamefont {de~la
  Cruz}, \citenamefont {Chi}, \citenamefont {Podlesnyak},\ and\ \citenamefont
  {Fisher}}]{PhysRevB.85.214434}%
  \BibitemOpen
  \bibfield  {author} {\bibinfo {author} {\bibfnamefont {M.~C.}\ \bibnamefont
  {Shapiro}}, \bibinfo {author} {\bibfnamefont {Scott~C.}\ \bibnamefont
  {Riggs}}, \bibinfo {author} {\bibfnamefont {M.~B.}\ \bibnamefont {Stone}},
  \bibinfo {author} {\bibfnamefont {C.~R.}\ \bibnamefont {de~la Cruz}},
  \bibinfo {author} {\bibfnamefont {S.}~\bibnamefont {Chi}}, \bibinfo {author}
  {\bibfnamefont {A.~A.}\ \bibnamefont {Podlesnyak}}, \ and\ \bibinfo {author}
  {\bibfnamefont {I.~R.}\ \bibnamefont {Fisher}},\ }\bibfield  {title}
  {\enquote {\bibinfo {title} {{Structure and magnetic properties of the
  pyrochlore iridate Y${}_{2}$Ir${}_{2}$O${}_{7}$}},}\ }\href {\doibase
  10.1103/PhysRevB.85.214434} {\bibfield  {journal} {\bibinfo  {journal} {Phys.
  Rev. B}\ }\textbf {\bibinfo {volume} {85}},\ \bibinfo {pages} {214434}
  (\bibinfo {year} {2012})}\BibitemShut {NoStop}%
\bibitem [{\citenamefont {Disseler}\ \emph {et~al.}(2012)\citenamefont
  {Disseler}, \citenamefont {Dhital}, \citenamefont {Amato}, \citenamefont
  {Giblin}, \citenamefont {de~la Cruz}, \citenamefont {Wilson},\ and\
  \citenamefont {Graf}}]{PhysRevB.86.014428}%
  \BibitemOpen
  \bibfield  {author} {\bibinfo {author} {\bibfnamefont {S.~M.}\ \bibnamefont
  {Disseler}}, \bibinfo {author} {\bibfnamefont {Chetan}\ \bibnamefont
  {Dhital}}, \bibinfo {author} {\bibfnamefont {A.}~\bibnamefont {Amato}},
  \bibinfo {author} {\bibfnamefont {S.~R.}\ \bibnamefont {Giblin}}, \bibinfo
  {author} {\bibfnamefont {Clarina}\ \bibnamefont {de~la Cruz}}, \bibinfo
  {author} {\bibfnamefont {Stephen~D.}\ \bibnamefont {Wilson}}, \ and\ \bibinfo
  {author} {\bibfnamefont {M.~J.}\ \bibnamefont {Graf}},\ }\bibfield  {title}
  {\enquote {\bibinfo {title} {{Magnetic order in the pyrochlore iridates
  ${A}_{2}$Ir${}_{2}$O${}_{7}$ ($A$ = Y, Yb)}},}\ }\href {\doibase
  10.1103/PhysRevB.86.014428} {\bibfield  {journal} {\bibinfo  {journal} {Phys.
  Rev. B}\ }\textbf {\bibinfo {volume} {86}},\ \bibinfo {pages} {014428}
  (\bibinfo {year} {2012})}\BibitemShut {NoStop}%
\bibitem [{\citenamefont {Tafti}\ \emph {et~al.}(2012)\citenamefont {Tafti},
  \citenamefont {Ishikawa}, \citenamefont {McCollam}, \citenamefont
  {Nakatsuji},\ and\ \citenamefont {Julian}}]{PhysRevB.85.205104}%
  \BibitemOpen
  \bibfield  {author} {\bibinfo {author} {\bibfnamefont {F.~F.}\ \bibnamefont
  {Tafti}}, \bibinfo {author} {\bibfnamefont {J.~J.}\ \bibnamefont {Ishikawa}},
  \bibinfo {author} {\bibfnamefont {A.}~\bibnamefont {McCollam}}, \bibinfo
  {author} {\bibfnamefont {S.}~\bibnamefont {Nakatsuji}}, \ and\ \bibinfo
  {author} {\bibfnamefont {S.~R.}\ \bibnamefont {Julian}},\ }\bibfield  {title}
  {\enquote {\bibinfo {title} {{Pressure-tuned insulator to metal transition in
  ${\mathbf{Eu}}_{\mathbf{2}}{\mathbf{Ir}}_{\mathbf{2}}{\mathbf{O}}_{\mathbf{7}}$}},}\
  }\href {\doibase 10.1103/PhysRevB.85.205104} {\bibfield  {journal} {\bibinfo
  {journal} {Phys. Rev. B}\ }\textbf {\bibinfo {volume} {85}},\ \bibinfo
  {pages} {205104} (\bibinfo {year} {2012})}\BibitemShut {NoStop}%
\bibitem [{\citenamefont {Sagayama}\ \emph {et~al.}(2013)\citenamefont
  {Sagayama}, \citenamefont {Uematsu}, \citenamefont {Arima}, \citenamefont
  {Sugimoto}, \citenamefont {Ishikawa}, \citenamefont {O'Farrell},\ and\
  \citenamefont {Nakatsuji}}]{PhysRevB.87.100403}%
  \BibitemOpen
  \bibfield  {author} {\bibinfo {author} {\bibfnamefont {H.}~\bibnamefont
  {Sagayama}}, \bibinfo {author} {\bibfnamefont {D.}~\bibnamefont {Uematsu}},
  \bibinfo {author} {\bibfnamefont {T.}~\bibnamefont {Arima}}, \bibinfo
  {author} {\bibfnamefont {K.}~\bibnamefont {Sugimoto}}, \bibinfo {author}
  {\bibfnamefont {J.~J.}\ \bibnamefont {Ishikawa}}, \bibinfo {author}
  {\bibfnamefont {E.}~\bibnamefont {O'Farrell}}, \ and\ \bibinfo {author}
  {\bibfnamefont {S.}~\bibnamefont {Nakatsuji}},\ }\bibfield  {title} {\enquote
  {\bibinfo {title} {{Determination of long-range all-in-all-out ordering of
  Ir${}^{4+}$ moments in a pyrochlore iridate Eu${}_{2}$Ir${}_{2}$O${}_{7}$ by
  resonant x-ray diffraction}},}\ }\href {\doibase 10.1103/PhysRevB.87.100403}
  {\bibfield  {journal} {\bibinfo  {journal} {Phys. Rev. B}\ }\textbf {\bibinfo
  {volume} {87}},\ \bibinfo {pages} {100403} (\bibinfo {year}
  {2013})}\BibitemShut {NoStop}%
\bibitem [{\citenamefont {Donnerer}\ \emph {et~al.}(2016)\citenamefont
  {Donnerer}, \citenamefont {Rahn}, \citenamefont {Sala}, \citenamefont {Vale},
  \citenamefont {Pincini}, \citenamefont {Strempfer}, \citenamefont {Krisch},
  \citenamefont {Prabhakaran}, \citenamefont {Boothroyd},\ and\ \citenamefont
  {McMorrow}}]{PhysRevLett.117.037201}%
  \BibitemOpen
  \bibfield  {author} {\bibinfo {author} {\bibfnamefont {C.}~\bibnamefont
  {Donnerer}}, \bibinfo {author} {\bibfnamefont {M.~C.}\ \bibnamefont {Rahn}},
  \bibinfo {author} {\bibfnamefont {M.~Moretti}\ \bibnamefont {Sala}}, \bibinfo
  {author} {\bibfnamefont {J.~G.}\ \bibnamefont {Vale}}, \bibinfo {author}
  {\bibfnamefont {D.}~\bibnamefont {Pincini}}, \bibinfo {author} {\bibfnamefont
  {J.}~\bibnamefont {Strempfer}}, \bibinfo {author} {\bibfnamefont
  {M.}~\bibnamefont {Krisch}}, \bibinfo {author} {\bibfnamefont
  {D.}~\bibnamefont {Prabhakaran}}, \bibinfo {author} {\bibfnamefont {A.~T.}\
  \bibnamefont {Boothroyd}}, \ and\ \bibinfo {author} {\bibfnamefont {D.~F.}\
  \bibnamefont {McMorrow}},\ }\bibfield  {title} {\enquote {\bibinfo {title}
  {{All-in--all-Out Magnetic Order and Propagating Spin Waves in
  ${\mathrm{Sm}}_{2}{\mathrm{Ir}}_{2}{\mathrm{O}}_{7}$}},}\ }\href {\doibase
  10.1103/PhysRevLett.117.037201} {\bibfield  {journal} {\bibinfo  {journal}
  {Phys. Rev. Lett.}\ }\textbf {\bibinfo {volume} {117}},\ \bibinfo {pages}
  {037201} (\bibinfo {year} {2016})}\BibitemShut {NoStop}%
\bibitem [{\citenamefont {Ueda}\ \emph {et~al.}(2015)\citenamefont {Ueda},
  \citenamefont {Fujioka}, \citenamefont {Yang}, \citenamefont {Shiogai},
  \citenamefont {Tsukazaki}, \citenamefont {Nakamura}, \citenamefont {Awaji},
  \citenamefont {Nagaosa},\ and\ \citenamefont
  {Tokura}}]{PhysRevLett.115.056402}%
  \BibitemOpen
  \bibfield  {author} {\bibinfo {author} {\bibfnamefont {K.}~\bibnamefont
  {Ueda}}, \bibinfo {author} {\bibfnamefont {J.}~\bibnamefont {Fujioka}},
  \bibinfo {author} {\bibfnamefont {B.-J.}\ \bibnamefont {Yang}}, \bibinfo
  {author} {\bibfnamefont {J.}~\bibnamefont {Shiogai}}, \bibinfo {author}
  {\bibfnamefont {A.}~\bibnamefont {Tsukazaki}}, \bibinfo {author}
  {\bibfnamefont {S.}~\bibnamefont {Nakamura}}, \bibinfo {author}
  {\bibfnamefont {S.}~\bibnamefont {Awaji}}, \bibinfo {author} {\bibfnamefont
  {N.}~\bibnamefont {Nagaosa}}, \ and\ \bibinfo {author} {\bibfnamefont
  {Y.}~\bibnamefont {Tokura}},\ }\bibfield  {title} {\enquote {\bibinfo {title}
  {{Magnetic Field-Induced Insulator-Semimetal Transition in a Pyrochlore
  ${\mathrm{Nd}}_{2}{\mathrm{Ir}}_{2}{\mathrm{O}}_{7}$}},}\ }\href {\doibase
  10.1103/PhysRevLett.115.056402} {\bibfield  {journal} {\bibinfo  {journal}
  {Phys. Rev. Lett.}\ }\textbf {\bibinfo {volume} {115}},\ \bibinfo {pages}
  {056402} (\bibinfo {year} {2015})}\BibitemShut {NoStop}%
\bibitem [{\citenamefont {Disseler}(2014)}]{PhysRevB.89.140413}%
  \BibitemOpen
  \bibfield  {author} {\bibinfo {author} {\bibfnamefont {Steven~M.}\
  \bibnamefont {Disseler}},\ }\bibfield  {title} {\enquote {\bibinfo {title}
  {{Direct evidence for the all-in/all-out magnetic structure in the pyrochlore
  iridates from muon spin relaxation}},}\ }\href {\doibase
  10.1103/PhysRevB.89.140413} {\bibfield  {journal} {\bibinfo  {journal} {Phys.
  Rev. B}\ }\textbf {\bibinfo {volume} {89}},\ \bibinfo {pages} {140413}
  (\bibinfo {year} {2014})}\BibitemShut {NoStop}%
\bibitem [{\citenamefont {Hozoi}\ \emph {et~al.}(2014)\citenamefont {Hozoi},
  \citenamefont {Gretarsson}, \citenamefont {Clancy}, \citenamefont {Jeon},
  \citenamefont {Lee}, \citenamefont {Kim}, \citenamefont {Yushankhai},
  \citenamefont {Fulde}, \citenamefont {Casa}, \citenamefont {Gog},
  \citenamefont {Kim}, \citenamefont {Said}, \citenamefont {Upton},
  \citenamefont {Kim},\ and\ \citenamefont {van~den
  Brink}}]{PhysRevB.89.115111}%
  \BibitemOpen
  \bibfield  {author} {\bibinfo {author} {\bibfnamefont {L.}~\bibnamefont
  {Hozoi}}, \bibinfo {author} {\bibfnamefont {H.}~\bibnamefont {Gretarsson}},
  \bibinfo {author} {\bibfnamefont {J.~P.}\ \bibnamefont {Clancy}}, \bibinfo
  {author} {\bibfnamefont {B.-G.}\ \bibnamefont {Jeon}}, \bibinfo {author}
  {\bibfnamefont {B.}~\bibnamefont {Lee}}, \bibinfo {author} {\bibfnamefont
  {K.~H.}\ \bibnamefont {Kim}}, \bibinfo {author} {\bibfnamefont
  {V.}~\bibnamefont {Yushankhai}}, \bibinfo {author} {\bibfnamefont {Peter}\
  \bibnamefont {Fulde}}, \bibinfo {author} {\bibfnamefont {D.}~\bibnamefont
  {Casa}}, \bibinfo {author} {\bibfnamefont {T.}~\bibnamefont {Gog}}, \bibinfo
  {author} {\bibfnamefont {Jungho}\ \bibnamefont {Kim}}, \bibinfo {author}
  {\bibfnamefont {A.~H.}\ \bibnamefont {Said}}, \bibinfo {author}
  {\bibfnamefont {M.~H.}\ \bibnamefont {Upton}}, \bibinfo {author}
  {\bibfnamefont {Young-June}\ \bibnamefont {Kim}}, \ and\ \bibinfo {author}
  {\bibfnamefont {Jeroen}\ \bibnamefont {van~den Brink}},\ }\bibfield  {title}
  {\enquote {\bibinfo {title} {{Longer-range lattice anisotropy strongly
  competing with spin-orbit interactions in pyrochlore iridates}},}\ }\href
  {\doibase 10.1103/PhysRevB.89.115111} {\bibfield  {journal} {\bibinfo
  {journal} {Phys. Rev. B}\ }\textbf {\bibinfo {volume} {89}},\ \bibinfo
  {pages} {115111} (\bibinfo {year} {2014})}\BibitemShut {NoStop}%
\bibitem [{\citenamefont {Nakayama}\ \emph {et~al.}(2016)\citenamefont
  {Nakayama}, \citenamefont {Kondo}, \citenamefont {Tian}, \citenamefont
  {Ishikawa}, \citenamefont {Halim}, \citenamefont {Bareille}, \citenamefont
  {Malaeb}, \citenamefont {Kuroda}, \citenamefont {Tomita}, \citenamefont
  {Ideta}, \citenamefont {Tanaka}, \citenamefont {Matsunami}, \citenamefont
  {Kimura}, \citenamefont {Inami}, \citenamefont {Ono}, \citenamefont
  {Kumigashira}, \citenamefont {Balents}, \citenamefont {Nakatsuji},\ and\
  \citenamefont {Shin}}]{PhysRevLett.117.056403}%
  \BibitemOpen
  \bibfield  {author} {\bibinfo {author} {\bibfnamefont {M.}~\bibnamefont
  {Nakayama}}, \bibinfo {author} {\bibfnamefont {Takeshi}\ \bibnamefont
  {Kondo}}, \bibinfo {author} {\bibfnamefont {Z.}~\bibnamefont {Tian}},
  \bibinfo {author} {\bibfnamefont {J.~J.}\ \bibnamefont {Ishikawa}}, \bibinfo
  {author} {\bibfnamefont {M.}~\bibnamefont {Halim}}, \bibinfo {author}
  {\bibfnamefont {C.}~\bibnamefont {Bareille}}, \bibinfo {author}
  {\bibfnamefont {W.}~\bibnamefont {Malaeb}}, \bibinfo {author} {\bibfnamefont
  {K.}~\bibnamefont {Kuroda}}, \bibinfo {author} {\bibfnamefont
  {T.}~\bibnamefont {Tomita}}, \bibinfo {author} {\bibfnamefont
  {S.}~\bibnamefont {Ideta}}, \bibinfo {author} {\bibfnamefont
  {K.}~\bibnamefont {Tanaka}}, \bibinfo {author} {\bibfnamefont
  {M.}~\bibnamefont {Matsunami}}, \bibinfo {author} {\bibfnamefont
  {S.}~\bibnamefont {Kimura}}, \bibinfo {author} {\bibfnamefont
  {N.}~\bibnamefont {Inami}}, \bibinfo {author} {\bibfnamefont
  {K.}~\bibnamefont {Ono}}, \bibinfo {author} {\bibfnamefont {H.}~\bibnamefont
  {Kumigashira}}, \bibinfo {author} {\bibfnamefont {L.}~\bibnamefont
  {Balents}}, \bibinfo {author} {\bibfnamefont {S.}~\bibnamefont {Nakatsuji}},
  \ and\ \bibinfo {author} {\bibfnamefont {S.}~\bibnamefont {Shin}},\
  }\bibfield  {title} {\enquote {\bibinfo {title} {{Slater to Mott Crossover in
  the Metal to Insulator Transition of
  ${\mathrm{Nd}}_{2}{\mathrm{Ir}}_{2}{\mathrm{O}}_{7}$}},}\ }\href {\doibase
  10.1103/PhysRevLett.117.056403} {\bibfield  {journal} {\bibinfo  {journal}
  {Phys. Rev. Lett.}\ }\textbf {\bibinfo {volume} {117}},\ \bibinfo {pages}
  {056403} (\bibinfo {year} {2016})}\BibitemShut {NoStop}%
\bibitem [{\citenamefont {Guo}\ \emph {et~al.}(2013)\citenamefont {Guo},
  \citenamefont {Matsuhira}, \citenamefont {Kawasaki}, \citenamefont
  {Wakeshima}, \citenamefont {Hinatsu}, \citenamefont {Watanabe},\ and\
  \citenamefont {Xu}}]{PhysRevB.88.060411}%
  \BibitemOpen
  \bibfield  {author} {\bibinfo {author} {\bibfnamefont {Hanjie}\ \bibnamefont
  {Guo}}, \bibinfo {author} {\bibfnamefont {Kazuyuki}\ \bibnamefont
  {Matsuhira}}, \bibinfo {author} {\bibfnamefont {Ikuto}\ \bibnamefont
  {Kawasaki}}, \bibinfo {author} {\bibfnamefont {Makoto}\ \bibnamefont
  {Wakeshima}}, \bibinfo {author} {\bibfnamefont {Yukio}\ \bibnamefont
  {Hinatsu}}, \bibinfo {author} {\bibfnamefont {Isao}\ \bibnamefont
  {Watanabe}}, \ and\ \bibinfo {author} {\bibfnamefont {Zhu-an}\ \bibnamefont
  {Xu}},\ }\bibfield  {title} {\enquote {\bibinfo {title} {{Magnetic order in
  the pyrochlore iridate Nd${}_{2}$Ir${}_{2}$O${}_{7}$ probed by muon spin
  relaxation}},}\ }\href {\doibase 10.1103/PhysRevB.88.060411} {\bibfield
  {journal} {\bibinfo  {journal} {Phys. Rev. B}\ }\textbf {\bibinfo {volume}
  {88}},\ \bibinfo {pages} {060411} (\bibinfo {year} {2013})}\BibitemShut
  {NoStop}%
\bibitem [{\citenamefont {Ueda}\ \emph {et~al.}(2014)\citenamefont {Ueda},
  \citenamefont {Fujioka}, \citenamefont {Takahashi}, \citenamefont {Suzuki},
  \citenamefont {Ishiwata}, \citenamefont {Taguchi}, \citenamefont {Kawasaki},\
  and\ \citenamefont {Tokura}}]{PhysRevB.89.075127}%
  \BibitemOpen
  \bibfield  {author} {\bibinfo {author} {\bibfnamefont {K.}~\bibnamefont
  {Ueda}}, \bibinfo {author} {\bibfnamefont {J.}~\bibnamefont {Fujioka}},
  \bibinfo {author} {\bibfnamefont {Y.}~\bibnamefont {Takahashi}}, \bibinfo
  {author} {\bibfnamefont {T.}~\bibnamefont {Suzuki}}, \bibinfo {author}
  {\bibfnamefont {S.}~\bibnamefont {Ishiwata}}, \bibinfo {author}
  {\bibfnamefont {Y.}~\bibnamefont {Taguchi}}, \bibinfo {author} {\bibfnamefont
  {M.}~\bibnamefont {Kawasaki}}, \ and\ \bibinfo {author} {\bibfnamefont
  {Y.}~\bibnamefont {Tokura}},\ }\bibfield  {title} {\enquote {\bibinfo {title}
  {{Anomalous domain-wall conductance in pyrochlore-type
  ${\mathrm{Nd}}_{2}{\mathrm{Ir}}_{2}{\mathrm{O}}_{7}$ on the verge of the
  metal-insulator transition}},}\ }\href {\doibase 10.1103/PhysRevB.89.075127}
  {\bibfield  {journal} {\bibinfo  {journal} {Phys. Rev. B}\ }\textbf {\bibinfo
  {volume} {89}},\ \bibinfo {pages} {075127} (\bibinfo {year}
  {2014})}\BibitemShut {NoStop}%
\bibitem [{\citenamefont {Lefran\ifmmode~\mbox{\c{c}}\else \c{c}\fi{}ois}\
  \emph {et~al.}(2015)\citenamefont {Lefran\ifmmode~\mbox{\c{c}}\else
  \c{c}\fi{}ois}, \citenamefont {Simonet}, \citenamefont {Ballou},
  \citenamefont {Lhotel}, \citenamefont {Hadj-Azzem}, \citenamefont
  {Kodjikian}, \citenamefont {Lejay}, \citenamefont {Manuel}, \citenamefont
  {Khalyavin},\ and\ \citenamefont {Chapon}}]{PhysRevLett.114.247202}%
  \BibitemOpen
  \bibfield  {author} {\bibinfo {author} {\bibfnamefont {E.}~\bibnamefont
  {Lefran\ifmmode~\mbox{\c{c}}\else \c{c}\fi{}ois}}, \bibinfo {author}
  {\bibfnamefont {V.}~\bibnamefont {Simonet}}, \bibinfo {author} {\bibfnamefont
  {R.}~\bibnamefont {Ballou}}, \bibinfo {author} {\bibfnamefont
  {E.}~\bibnamefont {Lhotel}}, \bibinfo {author} {\bibfnamefont
  {A.}~\bibnamefont {Hadj-Azzem}}, \bibinfo {author} {\bibfnamefont
  {S.}~\bibnamefont {Kodjikian}}, \bibinfo {author} {\bibfnamefont
  {P.}~\bibnamefont {Lejay}}, \bibinfo {author} {\bibfnamefont
  {P.}~\bibnamefont {Manuel}}, \bibinfo {author} {\bibfnamefont
  {D.}~\bibnamefont {Khalyavin}}, \ and\ \bibinfo {author} {\bibfnamefont
  {L.~C.}\ \bibnamefont {Chapon}},\ }\bibfield  {title} {\enquote {\bibinfo
  {title} {{Anisotropy-Tuned Magnetic Order in Pyrochlore Iridates}},}\ }\href
  {\doibase 10.1103/PhysRevLett.114.247202} {\bibfield  {journal} {\bibinfo
  {journal} {Phys. Rev. Lett.}\ }\textbf {\bibinfo {volume} {114}},\ \bibinfo
  {pages} {247202} (\bibinfo {year} {2015})}\BibitemShut {NoStop}%
\bibitem [{\citenamefont {Disseler}\ \emph {et~al.}(2013)\citenamefont
  {Disseler}, \citenamefont {Giblin}, \citenamefont {Dhital}, \citenamefont
  {Lukas}, \citenamefont {Wilson},\ and\ \citenamefont
  {Graf}}]{PhysRevB.87.060403}%
  \BibitemOpen
  \bibfield  {author} {\bibinfo {author} {\bibfnamefont {S.~M.}\ \bibnamefont
  {Disseler}}, \bibinfo {author} {\bibfnamefont {S.~R.}\ \bibnamefont
  {Giblin}}, \bibinfo {author} {\bibfnamefont {Chetan}\ \bibnamefont {Dhital}},
  \bibinfo {author} {\bibfnamefont {K.~C.}\ \bibnamefont {Lukas}}, \bibinfo
  {author} {\bibfnamefont {Stephen~D.}\ \bibnamefont {Wilson}}, \ and\ \bibinfo
  {author} {\bibfnamefont {M.~J.}\ \bibnamefont {Graf}},\ }\bibfield  {title}
  {\enquote {\bibinfo {title} {{Magnetization and Hall effect studies on the
  pyrochlore iridate Nd${}_{2}$Ir${}_{2}$O${}_{7}$}},}\ }\href {\doibase
  10.1103/PhysRevB.87.060403} {\bibfield  {journal} {\bibinfo  {journal} {Phys.
  Rev. B}\ }\textbf {\bibinfo {volume} {87}},\ \bibinfo {pages} {060403}
  (\bibinfo {year} {2013})}\BibitemShut {NoStop}%
\bibitem [{\citenamefont {Takatsu}\ \emph {et~al.}(2014)\citenamefont
  {Takatsu}, \citenamefont {Watanabe}, \citenamefont {Goto},\ and\
  \citenamefont {Kadowaki}}]{PhysRevB.90.235110}%
  \BibitemOpen
  \bibfield  {author} {\bibinfo {author} {\bibfnamefont {Hiroshi}\ \bibnamefont
  {Takatsu}}, \bibinfo {author} {\bibfnamefont {Kunihiko}\ \bibnamefont
  {Watanabe}}, \bibinfo {author} {\bibfnamefont {Kazuki}\ \bibnamefont {Goto}},
  \ and\ \bibinfo {author} {\bibfnamefont {Hiroaki}\ \bibnamefont {Kadowaki}},\
  }\bibfield  {title} {\enquote {\bibinfo {title} {{Comparative study of
  low-temperature x-ray diffraction experiments on
  ${R}_{2}{\mathrm{Ir}}_{2}{\mathrm{O}}_{7}$ ($R=\mathrm{Nd}$, Eu, and Pr)}},}\
  }\href {\doibase 10.1103/PhysRevB.90.235110} {\bibfield  {journal} {\bibinfo
  {journal} {Phys. Rev. B}\ }\textbf {\bibinfo {volume} {90}},\ \bibinfo
  {pages} {235110} (\bibinfo {year} {2014})}\BibitemShut {NoStop}%
\bibitem [{\citenamefont {Zhao}\ \emph {et~al.}(2011)\citenamefont {Zhao},
  \citenamefont {Mackie}, \citenamefont {MacLaughlin}, \citenamefont {Bernal},
  \citenamefont {Ishikawa}, \citenamefont {Ohta},\ and\ \citenamefont
  {Nakatsuji}}]{PhysRevB.83.180402}%
  \BibitemOpen
  \bibfield  {author} {\bibinfo {author} {\bibfnamefont {Songrui}\ \bibnamefont
  {Zhao}}, \bibinfo {author} {\bibfnamefont {J.~M.}\ \bibnamefont {Mackie}},
  \bibinfo {author} {\bibfnamefont {D.~E.}\ \bibnamefont {MacLaughlin}},
  \bibinfo {author} {\bibfnamefont {O.~O.}\ \bibnamefont {Bernal}}, \bibinfo
  {author} {\bibfnamefont {J.~J.}\ \bibnamefont {Ishikawa}}, \bibinfo {author}
  {\bibfnamefont {Y.}~\bibnamefont {Ohta}}, \ and\ \bibinfo {author}
  {\bibfnamefont {S.}~\bibnamefont {Nakatsuji}},\ }\bibfield  {title} {\enquote
  {\bibinfo {title} {{Magnetic transition, long-range order, and moment
  fluctuations in the pyrochlore iridate Eu${}_{2}$Ir${}_{2}$O${}_{7}$}},}\
  }\href {\doibase 10.1103/PhysRevB.83.180402} {\bibfield  {journal} {\bibinfo
  {journal} {Phys. Rev. B}\ }\textbf {\bibinfo {volume} {83}},\ \bibinfo
  {pages} {180402} (\bibinfo {year} {2011})}\BibitemShut {NoStop}%
\bibitem [{\citenamefont {Guo}\ \emph {et~al.}(2016)\citenamefont {Guo},
  \citenamefont {Ritter},\ and\ \citenamefont {Komarek}}]{PhysRevB.94.161102}%
  \BibitemOpen
  \bibfield  {author} {\bibinfo {author} {\bibfnamefont {H.}~\bibnamefont
  {Guo}}, \bibinfo {author} {\bibfnamefont {C.}~\bibnamefont {Ritter}}, \ and\
  \bibinfo {author} {\bibfnamefont {A.~C.}\ \bibnamefont {Komarek}},\
  }\bibfield  {title} {\enquote {\bibinfo {title} {{Direct determination of the
  spin structure of ${\mathrm{Nd}}_{2}{\mathrm{Ir}}_{2}{\mathrm{O}}_{7}$ by
  means of neutron diffraction}},}\ }\href {\doibase
  10.1103/PhysRevB.94.161102} {\bibfield  {journal} {\bibinfo  {journal} {Phys.
  Rev. B}\ }\textbf {\bibinfo {volume} {94}},\ \bibinfo {pages} {161102}
  (\bibinfo {year} {2016})}\BibitemShut {NoStop}%
\bibitem [{\citenamefont {Uematsu}\ \emph {et~al.}(2015)\citenamefont
  {Uematsu}, \citenamefont {Sagayama}, \citenamefont {Arima}, \citenamefont
  {Ishikawa}, \citenamefont {Nakatsuji}, \citenamefont {Takagi}, \citenamefont
  {Yoshida}, \citenamefont {Mizuki},\ and\ \citenamefont
  {Ishii}}]{PhysRevB.92.094405}%
  \BibitemOpen
  \bibfield  {author} {\bibinfo {author} {\bibfnamefont {Daisuke}\ \bibnamefont
  {Uematsu}}, \bibinfo {author} {\bibfnamefont {Hajime}\ \bibnamefont
  {Sagayama}}, \bibinfo {author} {\bibfnamefont {Taka-hisa}\ \bibnamefont
  {Arima}}, \bibinfo {author} {\bibfnamefont {Jun~J.}\ \bibnamefont
  {Ishikawa}}, \bibinfo {author} {\bibfnamefont {Satoru}\ \bibnamefont
  {Nakatsuji}}, \bibinfo {author} {\bibfnamefont {Hidenori}\ \bibnamefont
  {Takagi}}, \bibinfo {author} {\bibfnamefont {Masahiro}\ \bibnamefont
  {Yoshida}}, \bibinfo {author} {\bibfnamefont {Jun'ichiro}\ \bibnamefont
  {Mizuki}}, \ and\ \bibinfo {author} {\bibfnamefont {Kenji}\ \bibnamefont
  {Ishii}},\ }\bibfield  {title} {\enquote {\bibinfo {title} {{Large
  trigonal-field effect on spin-orbit coupled states in a pyrochlore
  iridate}},}\ }\href {\doibase 10.1103/PhysRevB.92.094405} {\bibfield
  {journal} {\bibinfo  {journal} {Phys. Rev. B}\ }\textbf {\bibinfo {volume}
  {92}},\ \bibinfo {pages} {094405} (\bibinfo {year} {2015})}\BibitemShut
  {NoStop}%
\bibitem [{\citenamefont {Yang}\ \emph {et~al.}(2017)\citenamefont {Yang},
  \citenamefont {Zhu}, \citenamefont {Zhou}, \citenamefont {Ling},
  \citenamefont {Choi}, \citenamefont {Lee}, \citenamefont {Losovyj},
  \citenamefont {Lu},\ and\ \citenamefont {Zhang}}]{PhysRevB.96.094437}%
  \BibitemOpen
  \bibfield  {author} {\bibinfo {author} {\bibfnamefont {W.~C.}\ \bibnamefont
  {Yang}}, \bibinfo {author} {\bibfnamefont {W.~K.}\ \bibnamefont {Zhu}},
  \bibinfo {author} {\bibfnamefont {H.~D.}\ \bibnamefont {Zhou}}, \bibinfo
  {author} {\bibfnamefont {L.}~\bibnamefont {Ling}}, \bibinfo {author}
  {\bibfnamefont {E.~S.}\ \bibnamefont {Choi}}, \bibinfo {author}
  {\bibfnamefont {M.}~\bibnamefont {Lee}}, \bibinfo {author} {\bibfnamefont
  {Y.}~\bibnamefont {Losovyj}}, \bibinfo {author} {\bibfnamefont {Chi-Ken}\
  \bibnamefont {Lu}}, \ and\ \bibinfo {author} {\bibfnamefont {S.~X.}\
  \bibnamefont {Zhang}},\ }\bibfield  {title} {\enquote {\bibinfo {title}
  {{Robust pinning of magnetic moments in pyrochlore iridates}},}\ }\href
  {\doibase 10.1103/PhysRevB.96.094437} {\bibfield  {journal} {\bibinfo
  {journal} {Phys. Rev. B}\ }\textbf {\bibinfo {volume} {96}},\ \bibinfo
  {pages} {094437} (\bibinfo {year} {2017})}\BibitemShut {NoStop}%
\bibitem [{\citenamefont {Fujita}\ \emph {et~al.}(2018)\citenamefont {Fujita},
  \citenamefont {Kozuka}, \citenamefont {Matsuno}, \citenamefont {Uchida},
  \citenamefont {Tsukazaki}, \citenamefont {Arima},\ and\ \citenamefont
  {Kawasaki}}]{PhysRevMaterials.2.011402}%
  \BibitemOpen
  \bibfield  {author} {\bibinfo {author} {\bibfnamefont {T.~C.}\ \bibnamefont
  {Fujita}}, \bibinfo {author} {\bibfnamefont {Y.}~\bibnamefont {Kozuka}},
  \bibinfo {author} {\bibfnamefont {J.}~\bibnamefont {Matsuno}}, \bibinfo
  {author} {\bibfnamefont {M.}~\bibnamefont {Uchida}}, \bibinfo {author}
  {\bibfnamefont {A.}~\bibnamefont {Tsukazaki}}, \bibinfo {author}
  {\bibfnamefont {T.}~\bibnamefont {Arima}}, \ and\ \bibinfo {author}
  {\bibfnamefont {M.}~\bibnamefont {Kawasaki}},\ }\bibfield  {title} {\enquote
  {\bibinfo {title} {{All-in-all-out magnetic domain inversion in
  ${\mathrm{Tb}}_{2}{\mathrm{Ir}}_{2}{\mathrm{O}}_{7}$ with molecular fields
  antiparallel to external fields}},}\ }\href {\doibase
  10.1103/PhysRevMaterials.2.011402} {\bibfield  {journal} {\bibinfo  {journal}
  {Phys. Rev. Materials}\ }\textbf {\bibinfo {volume} {2}},\ \bibinfo {pages}
  {011402} (\bibinfo {year} {2018})}\BibitemShut {NoStop}%
\bibitem [{\citenamefont {{Wang}}\ \emph {et~al.}(2020)\citenamefont {{Wang}},
  \citenamefont {{Xu}}, \citenamefont {{Rischau}}, \citenamefont {{Bachar}},
  \citenamefont {{Michon}}, \citenamefont {{Teyssier}}, \citenamefont {{Qiu}},
  \citenamefont {{Ohtsuki}}, \citenamefont {{Cheng}}, \citenamefont
  {{Armitage}}, \citenamefont {{Nakatsuji}},\ and\ \citenamefont {{van der
  Marel}}}]{2020arXiv200512768W}%
  \BibitemOpen
  \bibfield  {author} {\bibinfo {author} {\bibfnamefont {K.}~\bibnamefont
  {{Wang}}}, \bibinfo {author} {\bibfnamefont {B.}~\bibnamefont {{Xu}}},
  \bibinfo {author} {\bibfnamefont {C.~W.}\ \bibnamefont {{Rischau}}}, \bibinfo
  {author} {\bibfnamefont {N.}~\bibnamefont {{Bachar}}}, \bibinfo {author}
  {\bibfnamefont {B.}~\bibnamefont {{Michon}}}, \bibinfo {author}
  {\bibfnamefont {J.}~\bibnamefont {{Teyssier}}}, \bibinfo {author}
  {\bibfnamefont {Y.}~\bibnamefont {{Qiu}}}, \bibinfo {author} {\bibfnamefont
  {T.}~\bibnamefont {{Ohtsuki}}}, \bibinfo {author} {\bibfnamefont {Bing}\
  \bibnamefont {{Cheng}}}, \bibinfo {author} {\bibfnamefont {N.~P.}\
  \bibnamefont {{Armitage}}}, \bibinfo {author} {\bibfnamefont
  {S.}~\bibnamefont {{Nakatsuji}}}, \ and\ \bibinfo {author} {\bibfnamefont
  {D.}~\bibnamefont {{van der Marel}}},\ }\bibfield  {title} {\enquote
  {\bibinfo {title} {{Unconventional free charge in the correlated semimetal
  Nd$_2$Ir$_2$O$_7$}},}\ }\href@noop {} {\bibfield  {journal} {\bibinfo
  {journal} {Nature Physics}\ ,\ \bibinfo {eid} {arXiv:2005.12768}} (\bibinfo
  {year} {2020})},\ \Eprint {http://arxiv.org/abs/2005.12768} {arXiv:2005.12768
  [cond-mat.str-el]} \BibitemShut {NoStop}%
\bibitem [{\citenamefont {Krajewska}\ \emph {et~al.}(2020)\citenamefont
  {Krajewska}, \citenamefont {Takayama}, \citenamefont {Dinnebier},
  \citenamefont {Yaresko}, \citenamefont {Ishii}, \citenamefont {Isobe},\ and\
  \citenamefont {Takagi}}]{PhysRevB.101.121101}%
  \BibitemOpen
  \bibfield  {author} {\bibinfo {author} {\bibfnamefont {A.}~\bibnamefont
  {Krajewska}}, \bibinfo {author} {\bibfnamefont {T.}~\bibnamefont {Takayama}},
  \bibinfo {author} {\bibfnamefont {R.}~\bibnamefont {Dinnebier}}, \bibinfo
  {author} {\bibfnamefont {A.}~\bibnamefont {Yaresko}}, \bibinfo {author}
  {\bibfnamefont {K.}~\bibnamefont {Ishii}}, \bibinfo {author} {\bibfnamefont
  {M.}~\bibnamefont {Isobe}}, \ and\ \bibinfo {author} {\bibfnamefont
  {H.}~\bibnamefont {Takagi}},\ }\bibfield  {title} {\enquote {\bibinfo {title}
  {{Almost pure ${J}_{\text{eff}}=\frac{1}{2}$ Mott state of
  $\mathrm{In}{}_{2}\mathrm{Ir}{}_{2}\mathrm{O}{}_{7}$ in the limit of reduced
  intersite hopping}},}\ }\href {\doibase 10.1103/PhysRevB.101.121101}
  {\bibfield  {journal} {\bibinfo  {journal} {Phys. Rev. B}\ }\textbf {\bibinfo
  {volume} {101}},\ \bibinfo {pages} {121101} (\bibinfo {year}
  {2020})}\BibitemShut {NoStop}%
\bibitem [{\citenamefont {Guo}\ \emph {et~al.}(2017)\citenamefont {Guo},
  \citenamefont {Ritter},\ and\ \citenamefont {Komarek}}]{PhysRevB.96.144415}%
  \BibitemOpen
  \bibfield  {author} {\bibinfo {author} {\bibfnamefont {H.}~\bibnamefont
  {Guo}}, \bibinfo {author} {\bibfnamefont {C.}~\bibnamefont {Ritter}}, \ and\
  \bibinfo {author} {\bibfnamefont {A.~C.}\ \bibnamefont {Komarek}},\
  }\bibfield  {title} {\enquote {\bibinfo {title} {{Magnetic structure of
  ${\mathbf{Tb}}_{\mathbf{2}}{\mathbf{Ir}}_{\mathbf{2}}{\mathbf{O}}_{\mathbf{7}}$
  determined by powder neutron diffraction}},}\ }\href {\doibase
  10.1103/PhysRevB.96.144415} {\bibfield  {journal} {\bibinfo  {journal} {Phys.
  Rev. B}\ }\textbf {\bibinfo {volume} {96}},\ \bibinfo {pages} {144415}
  (\bibinfo {year} {2017})}\BibitemShut {NoStop}%
\bibitem [{\citenamefont {Nakatsuji}\ \emph {et~al.}(2006)\citenamefont
  {Nakatsuji}, \citenamefont {Machida}, \citenamefont {Maeno}, \citenamefont
  {Tayama}, \citenamefont {Sakakibara}, \citenamefont {Duijn}, \citenamefont
  {Balicas}, \citenamefont {Millican}, \citenamefont {Macaluso},\ and\
  \citenamefont {Chan}}]{PhysRevLett.96.087204}%
  \BibitemOpen
  \bibfield  {author} {\bibinfo {author} {\bibfnamefont {S.}~\bibnamefont
  {Nakatsuji}}, \bibinfo {author} {\bibfnamefont {Y.}~\bibnamefont {Machida}},
  \bibinfo {author} {\bibfnamefont {Y.}~\bibnamefont {Maeno}}, \bibinfo
  {author} {\bibfnamefont {T.}~\bibnamefont {Tayama}}, \bibinfo {author}
  {\bibfnamefont {T.}~\bibnamefont {Sakakibara}}, \bibinfo {author}
  {\bibfnamefont {J.~van}\ \bibnamefont {Duijn}}, \bibinfo {author}
  {\bibfnamefont {L.}~\bibnamefont {Balicas}}, \bibinfo {author} {\bibfnamefont
  {J.~N.}\ \bibnamefont {Millican}}, \bibinfo {author} {\bibfnamefont {R.~T.}\
  \bibnamefont {Macaluso}}, \ and\ \bibinfo {author} {\bibfnamefont {Julia~Y.}\
  \bibnamefont {Chan}},\ }\bibfield  {title} {\enquote {\bibinfo {title}
  {{Metallic Spin-Liquid Behavior of the Geometrically Frustrated Kondo Lattice
  ${\mathrm{Pr}}_{2}{\mathrm{Ir}}_{2}{\mathrm{O}}_{7}$}},}\ }\href {\doibase
  10.1103/PhysRevLett.96.087204} {\bibfield  {journal} {\bibinfo  {journal}
  {Phys. Rev. Lett.}\ }\textbf {\bibinfo {volume} {96}},\ \bibinfo {pages}
  {087204} (\bibinfo {year} {2006})}\BibitemShut {NoStop}%
\bibitem [{\citenamefont {Machida}\ \emph {et~al.}(2010)\citenamefont
  {Machida}, \citenamefont {Nakatsuji}, \citenamefont {Onoda}, \citenamefont
  {Tayama},\ and\ \citenamefont {Sakakibara}}]{Machida}%
  \BibitemOpen
  \bibfield  {author} {\bibinfo {author} {\bibfnamefont {Yo}~\bibnamefont
  {Machida}}, \bibinfo {author} {\bibfnamefont {Satoru}\ \bibnamefont
  {Nakatsuji}}, \bibinfo {author} {\bibfnamefont {Shigeki}\ \bibnamefont
  {Onoda}}, \bibinfo {author} {\bibfnamefont {Takashi}\ \bibnamefont {Tayama}},
  \ and\ \bibinfo {author} {\bibfnamefont {Toshiro}\ \bibnamefont
  {Sakakibara}},\ }\bibfield  {title} {\enquote {\bibinfo {title}
  {Time-reversal symmetry breaking and spontaneous hall effect without magnetic
  dipole order},}\ }\href {\doibase 10.1038/nature08680} {\bibfield  {journal}
  {\bibinfo  {journal} {Nature}\ }\textbf {\bibinfo {volume} {463}},\ \bibinfo
  {pages} {210--213} (\bibinfo {year} {2010})}\BibitemShut {NoStop}%
\bibitem [{\citenamefont {{Kondo}}\ \emph {et~al.}(2015)\citenamefont
  {{Kondo}}, \citenamefont {{Nakayama}}, \citenamefont {{Chen}}, \citenamefont
  {{Ishikawa}}, \citenamefont {{Moon}}, \citenamefont {{Yamamoto}},
  \citenamefont {{Ota}}, \citenamefont {{Malaeb}}, \citenamefont {{Kanai}},
  \citenamefont {{Nakashima}}, \citenamefont {{Ishida}}, \citenamefont
  {{Yoshida}}, \citenamefont {{Yamamoto}}, \citenamefont {{Matsunami}},
  \citenamefont {{Kimura}}, \citenamefont {{Inami}}, \citenamefont {{Ono}},
  \citenamefont {{Kumigashira}}, \citenamefont {{Nakatsuji}}, \citenamefont
  {{Balents}},\ and\ \citenamefont {{Shin}}}]{2015NatCo...610042K}%
  \BibitemOpen
  \bibfield  {author} {\bibinfo {author} {\bibfnamefont {Takeshi}\ \bibnamefont
  {{Kondo}}}, \bibinfo {author} {\bibfnamefont {M.}~\bibnamefont {{Nakayama}}},
  \bibinfo {author} {\bibfnamefont {R.}~\bibnamefont {{Chen}}}, \bibinfo
  {author} {\bibfnamefont {J.~J.}\ \bibnamefont {{Ishikawa}}}, \bibinfo
  {author} {\bibfnamefont {E.~G.}\ \bibnamefont {{Moon}}}, \bibinfo {author}
  {\bibfnamefont {T.}~\bibnamefont {{Yamamoto}}}, \bibinfo {author}
  {\bibfnamefont {Y.}~\bibnamefont {{Ota}}}, \bibinfo {author} {\bibfnamefont
  {W.}~\bibnamefont {{Malaeb}}}, \bibinfo {author} {\bibfnamefont
  {H.}~\bibnamefont {{Kanai}}}, \bibinfo {author} {\bibfnamefont
  {Y.}~\bibnamefont {{Nakashima}}}, \bibinfo {author} {\bibfnamefont
  {Y.}~\bibnamefont {{Ishida}}}, \bibinfo {author} {\bibfnamefont
  {R.}~\bibnamefont {{Yoshida}}}, \bibinfo {author} {\bibfnamefont
  {H.}~\bibnamefont {{Yamamoto}}}, \bibinfo {author} {\bibfnamefont
  {M.}~\bibnamefont {{Matsunami}}}, \bibinfo {author} {\bibfnamefont
  {S.}~\bibnamefont {{Kimura}}}, \bibinfo {author} {\bibfnamefont
  {N.}~\bibnamefont {{Inami}}}, \bibinfo {author} {\bibfnamefont
  {K.}~\bibnamefont {{Ono}}}, \bibinfo {author} {\bibfnamefont
  {H.}~\bibnamefont {{Kumigashira}}}, \bibinfo {author} {\bibfnamefont
  {S.}~\bibnamefont {{Nakatsuji}}}, \bibinfo {author} {\bibfnamefont
  {L.}~\bibnamefont {{Balents}}}, \ and\ \bibinfo {author} {\bibfnamefont
  {S.}~\bibnamefont {{Shin}}},\ }\bibfield  {title} {\enquote {\bibinfo {title}
  {{Quadratic Fermi node in a 3D strongly correlated semimetal}},}\ }\href
  {\doibase 10.1038/ncomms10042} {\bibfield  {journal} {\bibinfo  {journal}
  {Nature Communications}\ }\textbf {\bibinfo {volume} {6}},\ \bibinfo {eid}
  {10042} (\bibinfo {year} {2015})},\ \Eprint {http://arxiv.org/abs/1510.07977}
  {arXiv:1510.07977 [cond-mat.str-el]} \BibitemShut {NoStop}%
\bibitem [{\citenamefont {Savary}\ \emph {et~al.}(2014)\citenamefont {Savary},
  \citenamefont {Moon},\ and\ \citenamefont {Balents}}]{PhysRevX.4.041027}%
  \BibitemOpen
  \bibfield  {author} {\bibinfo {author} {\bibfnamefont {Lucile}\ \bibnamefont
  {Savary}}, \bibinfo {author} {\bibfnamefont {Eun-Gook}\ \bibnamefont {Moon}},
  \ and\ \bibinfo {author} {\bibfnamefont {Leon}\ \bibnamefont {Balents}},\
  }\bibfield  {title} {\enquote {\bibinfo {title} {{New Type of Quantum
  Criticality in the Pyrochlore Iridates}},}\ }\href {\doibase
  10.1103/PhysRevX.4.041027} {\bibfield  {journal} {\bibinfo  {journal} {Phys.
  Rev. X}\ }\textbf {\bibinfo {volume} {4}},\ \bibinfo {pages} {041027}
  (\bibinfo {year} {2014})}\BibitemShut {NoStop}%
\bibitem [{\citenamefont {Moon}\ \emph {et~al.}(2013)\citenamefont {Moon},
  \citenamefont {Xu}, \citenamefont {Kim},\ and\ \citenamefont
  {Balents}}]{PhysRevLett.111.206401}%
  \BibitemOpen
  \bibfield  {author} {\bibinfo {author} {\bibfnamefont {Eun-Gook}\
  \bibnamefont {Moon}}, \bibinfo {author} {\bibfnamefont {Cenke}\ \bibnamefont
  {Xu}}, \bibinfo {author} {\bibfnamefont {Yong~Baek}\ \bibnamefont {Kim}}, \
  and\ \bibinfo {author} {\bibfnamefont {Leon}\ \bibnamefont {Balents}},\
  }\bibfield  {title} {\enquote {\bibinfo {title} {{Non-Fermi-Liquid and
  Topological States with Strong Spin-Orbit Coupling}},}\ }\href {\doibase
  10.1103/PhysRevLett.111.206401} {\bibfield  {journal} {\bibinfo  {journal}
  {Phys. Rev. Lett.}\ }\textbf {\bibinfo {volume} {111}},\ \bibinfo {pages}
  {206401} (\bibinfo {year} {2013})}\BibitemShut {NoStop}%
\bibitem [{\citenamefont {Yao}\ and\ \citenamefont
  {Chen}(2018)}]{PhysRevX.8.041039}%
  \BibitemOpen
  \bibfield  {author} {\bibinfo {author} {\bibfnamefont {Xu-Ping}\ \bibnamefont
  {Yao}}\ and\ \bibinfo {author} {\bibfnamefont {Gang}\ \bibnamefont {Chen}},\
  }\bibfield  {title} {\enquote {\bibinfo {title}
  {{${\mathrm{Pr}}_{2}{\mathrm{Ir}}_{2}{\mathrm{O}}_{7}$: When Luttinger
  Semimetal Meets Melko-Hertog-Gingras Spin Ice State}},}\ }\href {\doibase
  10.1103/PhysRevX.8.041039} {\bibfield  {journal} {\bibinfo  {journal} {Phys.
  Rev. X}\ }\textbf {\bibinfo {volume} {8}},\ \bibinfo {pages} {041039}
  (\bibinfo {year} {2018})}\BibitemShut {NoStop}%
\bibitem [{\citenamefont {Chen}(2016)}]{PhysRevB.94.205107}%
  \BibitemOpen
  \bibfield  {author} {\bibinfo {author} {\bibfnamefont {Gang}\ \bibnamefont
  {Chen}},\ }\bibfield  {title} {\enquote {\bibinfo {title} {{``Magnetic
  monopole'' condensation of the pyrochlore ice U(1) quantum spin liquid:
  Application to ${\mathrm{Pr}}_{2}{\mathrm{Ir}}_{2}{\mathrm{O}}_{7}$ and
  ${\mathrm{Yb}}_{2}{\mathrm{Ti}}_{2}{\mathrm{O}}_{7}$}},}\ }\href {\doibase
  10.1103/PhysRevB.94.205107} {\bibfield  {journal} {\bibinfo  {journal} {Phys.
  Rev. B}\ }\textbf {\bibinfo {volume} {94}},\ \bibinfo {pages} {205107}
  (\bibinfo {year} {2016})}\BibitemShut {NoStop}%
\bibitem [{\citenamefont {{Ohtsuki}}\ \emph {et~al.}(2017)\citenamefont
  {{Ohtsuki}}, \citenamefont {{Tian}}, \citenamefont {{Endo}}, \citenamefont
  {{Halim}}, \citenamefont {{Katsumoto}}, \citenamefont {{Kohama}},
  \citenamefont {{Kindo}}, \citenamefont {{Nakatsuji}},\ and\ \citenamefont
  {{Lippmaa}}}]{2017arXiv171107813O}%
  \BibitemOpen
  \bibfield  {author} {\bibinfo {author} {\bibfnamefont {Takumi}\ \bibnamefont
  {{Ohtsuki}}}, \bibinfo {author} {\bibfnamefont {Zhaoming}\ \bibnamefont
  {{Tian}}}, \bibinfo {author} {\bibfnamefont {Akira}\ \bibnamefont {{Endo}}},
  \bibinfo {author} {\bibfnamefont {Mario}\ \bibnamefont {{Halim}}}, \bibinfo
  {author} {\bibfnamefont {Shingo}\ \bibnamefont {{Katsumoto}}}, \bibinfo
  {author} {\bibfnamefont {Yoshimitsu}\ \bibnamefont {{Kohama}}}, \bibinfo
  {author} {\bibfnamefont {Koichi}\ \bibnamefont {{Kindo}}}, \bibinfo {author}
  {\bibfnamefont {Satoru}\ \bibnamefont {{Nakatsuji}}}, \ and\ \bibinfo
  {author} {\bibfnamefont {Mikk}\ \bibnamefont {{Lippmaa}}},\ }\bibfield
  {title} {\enquote {\bibinfo {title} {{Spontaneous Hall effect induced by
  strain in Pr$_2$Ir$_2$O$_7$ epitaxial thin films}},}\ }\href@noop {}
  {\bibfield  {journal} {\bibinfo  {journal} {arXiv e-prints}\ ,\ \bibinfo
  {eid} {arXiv:1711.07813}} (\bibinfo {year} {2017})},\ \Eprint
  {http://arxiv.org/abs/1711.07813} {arXiv:1711.07813 [cond-mat.str-el]}
  \BibitemShut {NoStop}%
\bibitem [{\citenamefont {MacLaughlin}\ \emph {et~al.}(2015)\citenamefont
  {MacLaughlin}, \citenamefont {Bernal}, \citenamefont {Shu}, \citenamefont
  {Ishikawa}, \citenamefont {Matsumoto}, \citenamefont {Wen}, \citenamefont
  {Mourigal}, \citenamefont {Stock}, \citenamefont {Ehlers}, \citenamefont
  {Broholm}, \citenamefont {Machida}, \citenamefont {Kimura}, \citenamefont
  {Nakatsuji}, \citenamefont {Shimura},\ and\ \citenamefont
  {Sakakibara}}]{PhysRevB.92.054432}%
  \BibitemOpen
  \bibfield  {author} {\bibinfo {author} {\bibfnamefont {D.~E.}\ \bibnamefont
  {MacLaughlin}}, \bibinfo {author} {\bibfnamefont {O.~O.}\ \bibnamefont
  {Bernal}}, \bibinfo {author} {\bibfnamefont {Lei}\ \bibnamefont {Shu}},
  \bibinfo {author} {\bibfnamefont {Jun}\ \bibnamefont {Ishikawa}}, \bibinfo
  {author} {\bibfnamefont {Yosuke}\ \bibnamefont {Matsumoto}}, \bibinfo
  {author} {\bibfnamefont {J.-J.}\ \bibnamefont {Wen}}, \bibinfo {author}
  {\bibfnamefont {M.}~\bibnamefont {Mourigal}}, \bibinfo {author}
  {\bibfnamefont {C.}~\bibnamefont {Stock}}, \bibinfo {author} {\bibfnamefont
  {G.}~\bibnamefont {Ehlers}}, \bibinfo {author} {\bibfnamefont {C.~L.}\
  \bibnamefont {Broholm}}, \bibinfo {author} {\bibfnamefont {Yo}~\bibnamefont
  {Machida}}, \bibinfo {author} {\bibfnamefont {Kenta}\ \bibnamefont {Kimura}},
  \bibinfo {author} {\bibfnamefont {Satoru}\ \bibnamefont {Nakatsuji}},
  \bibinfo {author} {\bibfnamefont {Yasuyuki}\ \bibnamefont {Shimura}}, \ and\
  \bibinfo {author} {\bibfnamefont {Toshiro}\ \bibnamefont {Sakakibara}},\
  }\bibfield  {title} {\enquote {\bibinfo {title} {{Unstable spin-ice order in
  the stuffed metallic pyrochlore
  ${\mathrm{Pr}}_{2+x}{\mathrm{Ir}}_{2\ensuremath{-}x}{\mathrm{O}}_{7\ensuremath{-}\ensuremath{\delta}}$}},}\
  }\href {\doibase 10.1103/PhysRevB.92.054432} {\bibfield  {journal} {\bibinfo
  {journal} {Phys. Rev. B}\ }\textbf {\bibinfo {volume} {92}},\ \bibinfo
  {pages} {054432} (\bibinfo {year} {2015})}\BibitemShut {NoStop}%
\bibitem [{\citenamefont {Yang}\ and\ \citenamefont
  {Kim}(2010)}]{kim_distortion}%
  \BibitemOpen
  \bibfield  {author} {\bibinfo {author} {\bibfnamefont {Bohm-Jung}\
  \bibnamefont {Yang}}\ and\ \bibinfo {author} {\bibfnamefont {Yong~Baek}\
  \bibnamefont {Kim}},\ }\bibfield  {title} {\enquote {\bibinfo {title}
  {Topological insulators and metal-insulator transition in the pyrochlore
  iridates},}\ }\href {\doibase 10.1103/PhysRevB.82.085111} {\bibfield
  {journal} {\bibinfo  {journal} {Phys. Rev. B}\ }\textbf {\bibinfo {volume}
  {82}},\ \bibinfo {pages} {085111} (\bibinfo {year} {2010})}\BibitemShut
  {NoStop}%
\bibitem [{\citenamefont {Chen}\ and\ \citenamefont
  {Hermele}(2012)}]{PhysRevB.86.235129}%
  \BibitemOpen
  \bibfield  {author} {\bibinfo {author} {\bibfnamefont {Gang}\ \bibnamefont
  {Chen}}\ and\ \bibinfo {author} {\bibfnamefont {Michael}\ \bibnamefont
  {Hermele}},\ }\bibfield  {title} {\enquote {\bibinfo {title} {{Magnetic
  orders and topological phases from $f$-$d$ exchange in pyrochlore
  iridates}},}\ }\href {\doibase 10.1103/PhysRevB.86.235129} {\bibfield
  {journal} {\bibinfo  {journal} {Phys. Rev. B}\ }\textbf {\bibinfo {volume}
  {86}},\ \bibinfo {pages} {235129} (\bibinfo {year} {2012})}\BibitemShut
  {NoStop}%
\bibitem [{\citenamefont {Wang}\ \emph
  {et~al.}(2017{\natexlab{a}})\citenamefont {Wang}, \citenamefont {Go},\ and\
  \citenamefont {Millis}}]{PhysRevB.96.195158}%
  \BibitemOpen
  \bibfield  {author} {\bibinfo {author} {\bibfnamefont {Runzhi}\ \bibnamefont
  {Wang}}, \bibinfo {author} {\bibfnamefont {Ara}\ \bibnamefont {Go}}, \ and\
  \bibinfo {author} {\bibfnamefont {Andrew}\ \bibnamefont {Millis}},\
  }\bibfield  {title} {\enquote {\bibinfo {title} {{Weyl rings and enhanced
  susceptibilities in pyrochlore iridates:
  $k\ifmmode\cdot\else\textperiodcentered\fi{}p$ analysis of cluster dynamical
  mean-field theory results}},}\ }\href {\doibase 10.1103/PhysRevB.96.195158}
  {\bibfield  {journal} {\bibinfo  {journal} {Phys. Rev. B}\ }\textbf {\bibinfo
  {volume} {96}},\ \bibinfo {pages} {195158} (\bibinfo {year}
  {2017}{\natexlab{a}})}\BibitemShut {NoStop}%
\bibitem [{\citenamefont {Rhim}\ and\ \citenamefont
  {Kim}(2015)}]{PhysRevB.91.115124}%
  \BibitemOpen
  \bibfield  {author} {\bibinfo {author} {\bibfnamefont {Jun-Won}\ \bibnamefont
  {Rhim}}\ and\ \bibinfo {author} {\bibfnamefont {Yong~Baek}\ \bibnamefont
  {Kim}},\ }\bibfield  {title} {\enquote {\bibinfo {title} {Quantum
  oscillations in the luttinger model with quadratic band touching:
  Applications to pyrochlore iridates},}\ }\href {\doibase
  10.1103/PhysRevB.91.115124} {\bibfield  {journal} {\bibinfo  {journal} {Phys.
  Rev. B}\ }\textbf {\bibinfo {volume} {91}},\ \bibinfo {pages} {115124}
  (\bibinfo {year} {2015})}\BibitemShut {NoStop}%
\bibitem [{\citenamefont {Yang}\ and\ \citenamefont
  {Nagaosa}(2014)}]{PhysRevLett.112.246402}%
  \BibitemOpen
  \bibfield  {author} {\bibinfo {author} {\bibfnamefont {Bohm-Jung}\
  \bibnamefont {Yang}}\ and\ \bibinfo {author} {\bibfnamefont {Naoto}\
  \bibnamefont {Nagaosa}},\ }\bibfield  {title} {\enquote {\bibinfo {title}
  {{Emergent Topological Phenomena in Thin Films of Pyrochlore Iridates}},}\
  }\href {\doibase 10.1103/PhysRevLett.112.246402} {\bibfield  {journal}
  {\bibinfo  {journal} {Phys. Rev. Lett.}\ }\textbf {\bibinfo {volume} {112}},\
  \bibinfo {pages} {246402} (\bibinfo {year} {2014})}\BibitemShut {NoStop}%
\bibitem [{\citenamefont {Lee}\ \emph {et~al.}(2013)\citenamefont {Lee},
  \citenamefont {Bhattacharjee},\ and\ \citenamefont
  {Kim}}]{PhysRevB.87.214416}%
  \BibitemOpen
  \bibfield  {author} {\bibinfo {author} {\bibfnamefont {Eric Kin-Ho}\
  \bibnamefont {Lee}}, \bibinfo {author} {\bibfnamefont {Subhro}\ \bibnamefont
  {Bhattacharjee}}, \ and\ \bibinfo {author} {\bibfnamefont {Yong~Baek}\
  \bibnamefont {Kim}},\ }\bibfield  {title} {\enquote {\bibinfo {title}
  {{Magnetic excitation spectra in pyrochlore iridates}},}\ }\href {\doibase
  10.1103/PhysRevB.87.214416} {\bibfield  {journal} {\bibinfo  {journal} {Phys.
  Rev. B}\ }\textbf {\bibinfo {volume} {87}},\ \bibinfo {pages} {214416}
  (\bibinfo {year} {2013})}\BibitemShut {NoStop}%
\bibitem [{\citenamefont {Wang}\ \emph
  {et~al.}(2017{\natexlab{b}})\citenamefont {Wang}, \citenamefont {Go},\ and\
  \citenamefont {Millis}}]{PhysRevB.95.045133}%
  \BibitemOpen
  \bibfield  {author} {\bibinfo {author} {\bibfnamefont {Runzhi}\ \bibnamefont
  {Wang}}, \bibinfo {author} {\bibfnamefont {Ara}\ \bibnamefont {Go}}, \ and\
  \bibinfo {author} {\bibfnamefont {Andrew~J.}\ \bibnamefont {Millis}},\
  }\bibfield  {title} {\enquote {\bibinfo {title} {{Electron interactions,
  spin-orbit coupling, and intersite correlations in pyrochlore iridates}},}\
  }\href {\doibase 10.1103/PhysRevB.95.045133} {\bibfield  {journal} {\bibinfo
  {journal} {Phys. Rev. B}\ }\textbf {\bibinfo {volume} {95}},\ \bibinfo
  {pages} {045133} (\bibinfo {year} {2017}{\natexlab{b}})}\BibitemShut
  {NoStop}%
\bibitem [{\citenamefont {Maekawa}\ and\ \citenamefont {{\sl et
  al}}(2004)}]{tmo}%
  \BibitemOpen
  \bibfield  {author} {\bibinfo {author} {\bibfnamefont {S.}~\bibnamefont
  {Maekawa}}\ and\ \bibinfo {author} {\bibnamefont {{\sl et al}}},\ }\href@noop
  {} {\emph {\bibinfo {title} {Physics of transition metal oxides}}}\ (\bibinfo
   {publisher} {Springer},\ \bibinfo {year} {2004})\BibitemShut {NoStop}%
\bibitem [{\citenamefont {Ross}\ \emph {et~al.}(2011)\citenamefont {Ross},
  \citenamefont {Savary}, \citenamefont {Gaulin},\ and\ \citenamefont
  {Balents}}]{PhysRevX.1.021002}%
  \BibitemOpen
  \bibfield  {author} {\bibinfo {author} {\bibfnamefont {Kate~A.}\ \bibnamefont
  {Ross}}, \bibinfo {author} {\bibfnamefont {Lucile}\ \bibnamefont {Savary}},
  \bibinfo {author} {\bibfnamefont {Bruce~D.}\ \bibnamefont {Gaulin}}, \ and\
  \bibinfo {author} {\bibfnamefont {Leon}\ \bibnamefont {Balents}},\ }\bibfield
   {title} {\enquote {\bibinfo {title} {{Quantum Excitations in Quantum Spin
  Ice}},}\ }\href {\doibase 10.1103/PhysRevX.1.021002} {\bibfield  {journal}
  {\bibinfo  {journal} {Phys. Rev. X}\ }\textbf {\bibinfo {volume} {1}},\
  \bibinfo {pages} {021002} (\bibinfo {year} {2011})}\BibitemShut {NoStop}%
\bibitem [{\citenamefont {Yan}\ \emph {et~al.}(2017)\citenamefont {Yan},
  \citenamefont {Benton}, \citenamefont {Jaubert},\ and\ \citenamefont
  {Shannon}}]{PhysRevB.95.094422}%
  \BibitemOpen
  \bibfield  {author} {\bibinfo {author} {\bibfnamefont {Han}\ \bibnamefont
  {Yan}}, \bibinfo {author} {\bibfnamefont {Owen}\ \bibnamefont {Benton}},
  \bibinfo {author} {\bibfnamefont {Ludovic}\ \bibnamefont {Jaubert}}, \ and\
  \bibinfo {author} {\bibfnamefont {Nic}\ \bibnamefont {Shannon}},\ }\bibfield
  {title} {\enquote {\bibinfo {title} {{Theory of multiple-phase competition in
  pyrochlore magnets with anisotropic exchange with application to
  ${\mathrm{Yb}}_{2}{\mathrm{Ti}}_{2}{\mathrm{O}}_{7},
  {\mathrm{Er}}_{2}{\mathrm{Ti}}_{2}{\mathrm{O}}_{7}$, and
  ${\mathrm{Er}}_{2}{\mathrm{Sn}}_{2}{\mathrm{O}}_{7}$}},}\ }\href {\doibase
  10.1103/PhysRevB.95.094422} {\bibfield  {journal} {\bibinfo  {journal} {Phys.
  Rev. B}\ }\textbf {\bibinfo {volume} {95}},\ \bibinfo {pages} {094422}
  (\bibinfo {year} {2017})}\BibitemShut {NoStop}%
\bibitem [{\citenamefont {Elhajal}\ \emph {et~al.}(2005)\citenamefont
  {Elhajal}, \citenamefont {Canals}, \citenamefont {Sunyer},\ and\
  \citenamefont {Lacroix}}]{pyrochlore_DM}%
  \BibitemOpen
  \bibfield  {author} {\bibinfo {author} {\bibfnamefont {Maged}\ \bibnamefont
  {Elhajal}}, \bibinfo {author} {\bibfnamefont {Benjamin}\ \bibnamefont
  {Canals}}, \bibinfo {author} {\bibfnamefont {Raimon}\ \bibnamefont {Sunyer}},
  \ and\ \bibinfo {author} {\bibfnamefont {Claudine}\ \bibnamefont {Lacroix}},\
  }\bibfield  {title} {\enquote {\bibinfo {title} {{Ordering in the pyrochlore
  antiferromagnet due to Dzyaloshinsky-Moriya interactions}},}\ }\href
  {\doibase 10.1103/PhysRevB.71.094420} {\bibfield  {journal} {\bibinfo
  {journal} {Phys. Rev. B}\ }\textbf {\bibinfo {volume} {71}},\ \bibinfo
  {pages} {094420} (\bibinfo {year} {2005})}\BibitemShut {NoStop}%
\bibitem [{\citenamefont {Savary}\ \emph {et~al.}(2012)\citenamefont {Savary},
  \citenamefont {Ross}, \citenamefont {Gaulin}, \citenamefont {Ruff},\ and\
  \citenamefont {Balents}}]{PhysRevLett.109.167201}%
  \BibitemOpen
  \bibfield  {author} {\bibinfo {author} {\bibfnamefont {Lucile}\ \bibnamefont
  {Savary}}, \bibinfo {author} {\bibfnamefont {Kate~A.}\ \bibnamefont {Ross}},
  \bibinfo {author} {\bibfnamefont {Bruce~D.}\ \bibnamefont {Gaulin}}, \bibinfo
  {author} {\bibfnamefont {Jacob P.~C.}\ \bibnamefont {Ruff}}, \ and\ \bibinfo
  {author} {\bibfnamefont {Leon}\ \bibnamefont {Balents}},\ }\bibfield  {title}
  {\enquote {\bibinfo {title} {{Order by Quantum Disorder in
  ${\mathrm{Er}}_{2}{\mathrm{Ti}}_{2}{\mathbf{O}}_{7}$}},}\ }\href {\doibase
  10.1103/PhysRevLett.109.167201} {\bibfield  {journal} {\bibinfo  {journal}
  {Phys. Rev. Lett.}\ }\textbf {\bibinfo {volume} {109}},\ \bibinfo {pages}
  {167201} (\bibinfo {year} {2012})}\BibitemShut {NoStop}%
\bibitem [{\citenamefont {Zener}(1951)}]{PhysRev.82.403}%
  \BibitemOpen
  \bibfield  {author} {\bibinfo {author} {\bibfnamefont {Clarence}\
  \bibnamefont {Zener}},\ }\bibfield  {title} {\enquote {\bibinfo {title}
  {{Interaction between the $d$-Shells in the Transition Metals. II.
  Ferromagnetic Compounds of Manganese with Perovskite Structure}},}\ }\href
  {\doibase 10.1103/PhysRev.82.403} {\bibfield  {journal} {\bibinfo  {journal}
  {Phys. Rev.}\ }\textbf {\bibinfo {volume} {82}},\ \bibinfo {pages} {403--405}
  (\bibinfo {year} {1951})}\BibitemShut {NoStop}%
\bibitem [{\citenamefont {Okuma}\ \emph {et~al.}(2020)\citenamefont {Okuma},
  \citenamefont {Ueta}, \citenamefont {Kuniyoshi}, \citenamefont {Fujisawa},
  \citenamefont {Smith}, \citenamefont {Hsu}, \citenamefont {Inagaki},
  \citenamefont {Si}, \citenamefont {Kawae}, \citenamefont {Lin}, \citenamefont
  {Chuang}, \citenamefont {Masuda}, \citenamefont {Kobayashi},\ and\
  \citenamefont {Okada}}]{okuma2020fermionic}%
  \BibitemOpen
  \bibfield  {author} {\bibinfo {author} {\bibfnamefont {R.}~\bibnamefont
  {Okuma}}, \bibinfo {author} {\bibfnamefont {D.}~\bibnamefont {Ueta}},
  \bibinfo {author} {\bibfnamefont {S.}~\bibnamefont {Kuniyoshi}}, \bibinfo
  {author} {\bibfnamefont {Y.}~\bibnamefont {Fujisawa}}, \bibinfo {author}
  {\bibfnamefont {B.}~\bibnamefont {Smith}}, \bibinfo {author} {\bibfnamefont
  {C.~H.}\ \bibnamefont {Hsu}}, \bibinfo {author} {\bibfnamefont
  {Y.}~\bibnamefont {Inagaki}}, \bibinfo {author} {\bibfnamefont
  {W.}~\bibnamefont {Si}}, \bibinfo {author} {\bibfnamefont {T.}~\bibnamefont
  {Kawae}}, \bibinfo {author} {\bibfnamefont {H.}~\bibnamefont {Lin}}, \bibinfo
  {author} {\bibfnamefont {F.~C.}\ \bibnamefont {Chuang}}, \bibinfo {author}
  {\bibfnamefont {T.}~\bibnamefont {Masuda}}, \bibinfo {author} {\bibfnamefont
  {R.}~\bibnamefont {Kobayashi}}, \ and\ \bibinfo {author} {\bibfnamefont
  {Y.}~\bibnamefont {Okada}},\ }\href@noop {} {\enquote {\bibinfo {title}
  {{Fermionic Order by Disorder in a van der Waals Antiferromagnet}},}\ }
  (\bibinfo {year} {2020}),\ \Eprint {http://arxiv.org/abs/2007.15193}
  {arXiv:2007.15193 [cond-mat.str-el]} \BibitemShut {NoStop}%
\bibitem [{\citenamefont {Zhou}\ \emph {et~al.}(2016)\citenamefont {Zhou},
  \citenamefont {Li}, \citenamefont {Waugh}, \citenamefont {Parham},
  \citenamefont {Kim}, \citenamefont {Sears}, \citenamefont {Gomes},
  \citenamefont {Kee}, \citenamefont {Kim},\ and\ \citenamefont
  {Dessau}}]{PhysRevB.94.161106}%
  \BibitemOpen
  \bibfield  {author} {\bibinfo {author} {\bibfnamefont {Xiaoqing}\
  \bibnamefont {Zhou}}, \bibinfo {author} {\bibfnamefont {Haoxiang}\
  \bibnamefont {Li}}, \bibinfo {author} {\bibfnamefont {J.~A.}\ \bibnamefont
  {Waugh}}, \bibinfo {author} {\bibfnamefont {S.}~\bibnamefont {Parham}},
  \bibinfo {author} {\bibfnamefont {Heung-Sik}\ \bibnamefont {Kim}}, \bibinfo
  {author} {\bibfnamefont {J.~A.}\ \bibnamefont {Sears}}, \bibinfo {author}
  {\bibfnamefont {A.}~\bibnamefont {Gomes}}, \bibinfo {author} {\bibfnamefont
  {Hae-Young}\ \bibnamefont {Kee}}, \bibinfo {author} {\bibfnamefont
  {Young-June}\ \bibnamefont {Kim}}, \ and\ \bibinfo {author} {\bibfnamefont
  {D.~S.}\ \bibnamefont {Dessau}},\ }\bibfield  {title} {\enquote {\bibinfo
  {title} {{Angle-resolved photoemission study of the Kitaev candidate
  $\ensuremath{\alpha}\ensuremath{-}{\mathrm{RuCl}}_{3}$}},}\ }\href {\doibase
  10.1103/PhysRevB.94.161106} {\bibfield  {journal} {\bibinfo  {journal} {Phys.
  Rev. B}\ }\textbf {\bibinfo {volume} {94}},\ \bibinfo {pages} {161106}
  (\bibinfo {year} {2016})}\BibitemShut {NoStop}%
\bibitem [{\citenamefont {Plumb}\ \emph {et~al.}(2014)\citenamefont {Plumb},
  \citenamefont {Clancy}, \citenamefont {Sandilands}, \citenamefont {Shankar},
  \citenamefont {Hu}, \citenamefont {Burch}, \citenamefont {Kee},\ and\
  \citenamefont {Kim}}]{PhysRevB.90.041112}%
  \BibitemOpen
  \bibfield  {author} {\bibinfo {author} {\bibfnamefont {K.~W.}\ \bibnamefont
  {Plumb}}, \bibinfo {author} {\bibfnamefont {J.~P.}\ \bibnamefont {Clancy}},
  \bibinfo {author} {\bibfnamefont {L.~J.}\ \bibnamefont {Sandilands}},
  \bibinfo {author} {\bibfnamefont {V.~Vijay}\ \bibnamefont {Shankar}},
  \bibinfo {author} {\bibfnamefont {Y.~F.}\ \bibnamefont {Hu}}, \bibinfo
  {author} {\bibfnamefont {K.~S.}\ \bibnamefont {Burch}}, \bibinfo {author}
  {\bibfnamefont {Hae-Young}\ \bibnamefont {Kee}}, \ and\ \bibinfo {author}
  {\bibfnamefont {Young-June}\ \bibnamefont {Kim}},\ }\bibfield  {title}
  {\enquote {\bibinfo {title}
  {{$\ensuremath{\alpha}\ensuremath{-}{\mathrm{RuCl}}_{3}$: A spin-orbit
  assisted Mott insulator on a honeycomb lattice}},}\ }\href {\doibase
  10.1103/PhysRevB.90.041112} {\bibfield  {journal} {\bibinfo  {journal} {Phys.
  Rev. B}\ }\textbf {\bibinfo {volume} {90}},\ \bibinfo {pages} {041112}
  (\bibinfo {year} {2014})}\BibitemShut {NoStop}%
\bibitem [{\citenamefont {Sandilands}\ \emph {et~al.}(2016)\citenamefont
  {Sandilands}, \citenamefont {Tian}, \citenamefont {Reijnders}, \citenamefont
  {Kim}, \citenamefont {Plumb}, \citenamefont {Kim}, \citenamefont {Kee},\ and\
  \citenamefont {Burch}}]{PhysRevB.93.075144}%
  \BibitemOpen
  \bibfield  {author} {\bibinfo {author} {\bibfnamefont {Luke~J.}\ \bibnamefont
  {Sandilands}}, \bibinfo {author} {\bibfnamefont {Yao}\ \bibnamefont {Tian}},
  \bibinfo {author} {\bibfnamefont {Anjan~A.}\ \bibnamefont {Reijnders}},
  \bibinfo {author} {\bibfnamefont {Heung-Sik}\ \bibnamefont {Kim}}, \bibinfo
  {author} {\bibfnamefont {K.~W.}\ \bibnamefont {Plumb}}, \bibinfo {author}
  {\bibfnamefont {Young-June}\ \bibnamefont {Kim}}, \bibinfo {author}
  {\bibfnamefont {Hae-Young}\ \bibnamefont {Kee}}, \ and\ \bibinfo {author}
  {\bibfnamefont {Kenneth~S.}\ \bibnamefont {Burch}},\ }\bibfield  {title}
  {\enquote {\bibinfo {title} {{Spin-orbit excitations and electronic structure
  of the putative Kitaev magnet
  $\ensuremath{\alpha}\ensuremath{-}{\mathrm{RuCl}}_{3}$}},}\ }\href {\doibase
  10.1103/PhysRevB.93.075144} {\bibfield  {journal} {\bibinfo  {journal} {Phys.
  Rev. B}\ }\textbf {\bibinfo {volume} {93}},\ \bibinfo {pages} {075144}
  (\bibinfo {year} {2016})}\BibitemShut {NoStop}%
\bibitem [{\citenamefont {Koitzsch}\ \emph {et~al.}(2016)\citenamefont
  {Koitzsch}, \citenamefont {Habenicht}, \citenamefont {M\"uller},
  \citenamefont {Knupfer}, \citenamefont {B\"uchner}, \citenamefont {Kandpal},
  \citenamefont {van~den Brink}, \citenamefont {Nowak}, \citenamefont
  {Isaeva},\ and\ \citenamefont {Doert}}]{PhysRevLett.117.126403}%
  \BibitemOpen
  \bibfield  {author} {\bibinfo {author} {\bibfnamefont {A.}~\bibnamefont
  {Koitzsch}}, \bibinfo {author} {\bibfnamefont {C.}~\bibnamefont {Habenicht}},
  \bibinfo {author} {\bibfnamefont {E.}~\bibnamefont {M\"uller}}, \bibinfo
  {author} {\bibfnamefont {M.}~\bibnamefont {Knupfer}}, \bibinfo {author}
  {\bibfnamefont {B.}~\bibnamefont {B\"uchner}}, \bibinfo {author}
  {\bibfnamefont {H.~C.}\ \bibnamefont {Kandpal}}, \bibinfo {author}
  {\bibfnamefont {J.}~\bibnamefont {van~den Brink}}, \bibinfo {author}
  {\bibfnamefont {D.}~\bibnamefont {Nowak}}, \bibinfo {author} {\bibfnamefont
  {A.}~\bibnamefont {Isaeva}}, \ and\ \bibinfo {author} {\bibfnamefont {Th.}\
  \bibnamefont {Doert}},\ }\bibfield  {title} {\enquote {\bibinfo {title}
  {{${J}_{\mathrm{eff}}$ Description of the Honeycomb Mott Insulator
  $\ensuremath{\alpha}\text{\ensuremath{-}}{\mathrm{RuCl}}_{3}$}},}\ }\href
  {\doibase 10.1103/PhysRevLett.117.126403} {\bibfield  {journal} {\bibinfo
  {journal} {Phys. Rev. Lett.}\ }\textbf {\bibinfo {volume} {117}},\ \bibinfo
  {pages} {126403} (\bibinfo {year} {2016})}\BibitemShut {NoStop}%
\bibitem [{\citenamefont {Ong}(2020)}]{Ong2020}%
  \BibitemOpen
  \bibfield  {author} {\bibinfo {author} {\bibfnamefont {P}~\bibnamefont
  {Ong}},\ }\href@noop {} {} (\bibinfo {year} {2020}),\ \bibinfo {note}
  {lecture at Princeton condensed matter summer school}\BibitemShut {NoStop}%
\bibitem [{\citenamefont {{Yokoi}}\ \emph {et~al.}(2020)\citenamefont
  {{Yokoi}}, \citenamefont {{Ma}}, \citenamefont {{Kasahara}}, \citenamefont
  {{Kasahara}}, \citenamefont {{Shibauchi}}, \citenamefont {{Kurita}},
  \citenamefont {{Tanaka}}, \citenamefont {{Nasu}}, \citenamefont {{Motome}},
  \citenamefont {{Hickey}}, \citenamefont {{Trebst}},\ and\ \citenamefont
  {{Matsuda}}}]{2020arXiv200101899Y}%
  \BibitemOpen
  \bibfield  {author} {\bibinfo {author} {\bibfnamefont {T.}~\bibnamefont
  {{Yokoi}}}, \bibinfo {author} {\bibfnamefont {S.}~\bibnamefont {{Ma}}},
  \bibinfo {author} {\bibfnamefont {Y.}~\bibnamefont {{Kasahara}}}, \bibinfo
  {author} {\bibfnamefont {S.}~\bibnamefont {{Kasahara}}}, \bibinfo {author}
  {\bibfnamefont {T.}~\bibnamefont {{Shibauchi}}}, \bibinfo {author}
  {\bibfnamefont {N.}~\bibnamefont {{Kurita}}}, \bibinfo {author}
  {\bibfnamefont {H.}~\bibnamefont {{Tanaka}}}, \bibinfo {author}
  {\bibfnamefont {J.}~\bibnamefont {{Nasu}}}, \bibinfo {author} {\bibfnamefont
  {Y.}~\bibnamefont {{Motome}}}, \bibinfo {author} {\bibfnamefont
  {C.}~\bibnamefont {{Hickey}}}, \bibinfo {author} {\bibfnamefont
  {S.}~\bibnamefont {{Trebst}}}, \ and\ \bibinfo {author} {\bibfnamefont
  {Y.}~\bibnamefont {{Matsuda}}},\ }\bibfield  {title} {\enquote {\bibinfo
  {title} {{Half-integer quantized anomalous thermal Hall effect in the Kitaev
  material $\alpha$-RuCl$_3$}},}\ }\href@noop {} {\bibfield  {journal}
  {\bibinfo  {journal} {arXiv e-prints}\ ,\ \bibinfo {eid} {arXiv:2001.01899}}
  (\bibinfo {year} {2020})},\ \Eprint {http://arxiv.org/abs/2001.01899}
  {arXiv:2001.01899 [cond-mat.str-el]} \BibitemShut {NoStop}%
\bibitem [{\citenamefont {Kasahara}\ \emph {et~al.}(2018)\citenamefont
  {Kasahara}, \citenamefont {Sugii}, \citenamefont {Ohnishi}, \citenamefont
  {Shimozawa}, \citenamefont {Yamashita}, \citenamefont {Kurita}, \citenamefont
  {Tanaka}, \citenamefont {Nasu}, \citenamefont {Motome}, \citenamefont
  {Shibauchi},\ and\ \citenamefont {Matsuda}}]{PhysRevLett.120.217205}%
  \BibitemOpen
  \bibfield  {author} {\bibinfo {author} {\bibfnamefont {Y.}~\bibnamefont
  {Kasahara}}, \bibinfo {author} {\bibfnamefont {K.}~\bibnamefont {Sugii}},
  \bibinfo {author} {\bibfnamefont {T.}~\bibnamefont {Ohnishi}}, \bibinfo
  {author} {\bibfnamefont {M.}~\bibnamefont {Shimozawa}}, \bibinfo {author}
  {\bibfnamefont {M.}~\bibnamefont {Yamashita}}, \bibinfo {author}
  {\bibfnamefont {N.}~\bibnamefont {Kurita}}, \bibinfo {author} {\bibfnamefont
  {H.}~\bibnamefont {Tanaka}}, \bibinfo {author} {\bibfnamefont
  {J.}~\bibnamefont {Nasu}}, \bibinfo {author} {\bibfnamefont {Y.}~\bibnamefont
  {Motome}}, \bibinfo {author} {\bibfnamefont {T.}~\bibnamefont {Shibauchi}}, \
  and\ \bibinfo {author} {\bibfnamefont {Y.}~\bibnamefont {Matsuda}},\
  }\bibfield  {title} {\enquote {\bibinfo {title} {{Unusual Thermal Hall Effect
  in a Kitaev Spin Liquid Candidate
  $\ensuremath{\alpha}\text{\ensuremath{-}}{\mathrm{RuCl}}_{3}$}},}\ }\href
  {\doibase 10.1103/PhysRevLett.120.217205} {\bibfield  {journal} {\bibinfo
  {journal} {Phys. Rev. Lett.}\ }\textbf {\bibinfo {volume} {120}},\ \bibinfo
  {pages} {217205} (\bibinfo {year} {2018})}\BibitemShut {NoStop}%
\end{thebibliography}%

\end{document}